%% file: covariance.tex
\documentclass[12pt]{article}
\usepackage{graphicx}
\usepackage[font=small,labelfont=bf]{caption}
\usepackage{diagbox}
\usepackage{url} 

\usepackage[nodisplayskipstretch]{setspace}
\newcommand{\blind}{0}

\usepackage[backend=biber, 
            style=apa,
            citestyle=authoryear-comp,
            natbib=true,
            url=false,
            doi=false,
            eprint=false]{biblatex}
\input{ST-input2}

\begin{document}




\if0\blind
{
  \title{Detecting changes in covariance via random matrix theory}
  \author{Sean Ryan and 
    Rebecca Killick 
\thanks{
    The authors gratefully acknowledge funding from EP/L015692/1 and NE/T012307/1.}\hspace{.2cm}\\
}
  \maketitle
} \fi

\if1\blind
{
  \bigskip
  \bigskip
  \bigskip
  \begin{center}
    {\LARGE\bf Detecting changes in covariance via random matrix theory}
\end{center}
  \medskip
} \fi

\begin{abstract} 
     A novel method is proposed for detecting changes in the covariance structure of moderate dimensional time series.
     This non-linear test statistic has a number of useful properties.
     Most importantly, it is independent of the underlying structure of the covariance matrix.
     We discuss how results from Random Matrix Theory,
     can be used to study the behaviour of our test statistic in a moderate dimensional setting
     (i.e. the number of variables is comparable to the length of the data).
     In particular, we demonstrate that the test statistic converges point wise to a normal
     distribution under the null hypothesis.
     We evaluate the performance of the proposed approach on a range of simulated datasets and
     find that it 
     outperforms a range of alternative recently proposed methods.
     Finally, we use our approach to study changes in the amount of  water on the surface of a plot of soil which feeds into model development for degradation of surface piping.
\end{abstract}

\noindent%
{\it Keywords:}  changepoint, ratio matrices, eigenvalue
\vfill

\newpage
\section{Introduction}
There is considerable infrastructure within the oil and gas industry that is on the surface of the ground.  
These are exposed to changing weather conditions and models are used to estimate the degradation of the assets and direct monitoring.  
These degradation models rely on estimates of the conditions that the assets are under.  
As the surrounding soils evolve, so do the absorption properties of the soil.  
This may result in assets sitting in waterlogged soil when the models are assuming that water drains away.  
Thus estimating how these soil properties vary over time, specifically regarding water absorption, is an important task.
Soil scientists typically measure water absorption through ground sensors at an individual location.  
This is not desirable as there may be large areas where assets are located.  
Recent research has explored video capture to estimate changes in surface water as a surrogate for absorption.  
Ambient lighting due to time of day and cloud cover hamper traditional mean-based methods for identifying surface water.  
In this paper, we develop a method for identifying changes within an image sequence (viewed as multivariate time series) that are linked to the changing presence of surface water. 

One approach to modelling changing behaviour in data  
is to assume that the changes occur at a small number of discrete time points known as changepoints.
Between changes, the data can be modelled as a set of stationary segments 
that satisfy standard modelling assumptions.
Changepoint methods are relevant in a wide range of applications including genetics~\citep{hocking2013learning}, network traffic analysis~\citep{rubin2016anomaly} and oceanography~\citep{tc-11-2149-2017}. 
We consider the specific case where the covariance structure of the data changes at each changepoint.  
While our focus here is on a specific application, the problem has wide applicability.  
For example,~\citet{stoehr2019detecting} examine changes in the covariance structure of functional Magnetic Resonance Imaging (fMRI) data, where a failure to satisfy stationarity assumptions can significantly contaminate any analysis, while  
~\citet{Wied2013} and~\citet{BERENS2015135} examine how changes in the covariance  of financial data can be used to improve stock portfolio optimisation.

The changepoint problem has a long history in the statistical literature, 
and contains two distinct but related problems, 
depending on whether the data is observed sequentially (online setting) or as a single batch (offline setting).
We focus on the latter and direct readers interested in the former to~\citet{tartakovsky2014sequential} for a thorough review.
In the univariate setting there is a vast literature on different methods for estimating 
changepoints, and there are a number of 
state of the art methods  
\citep{killick2012optimal,frick2014multiscale,fryzlewicz2014wild,maidstone2017optimal}.

The literature on detecting changes in multivariate time series has grown substantially in the last few years. In particular, many authors consider changes in the moderate dimensional setting, that is, where the number of the parameters of the model, is of the order of the number of data points. Much of this work considers changes in expectation where the series are uncorrelated \citep{grundy2020high,horvath2012change}. Furthermore a number of authors have examined detecting changes in expectation where only a subset of variables under observation change \citep{enikeeva2019,jirak2015uniform,wang2016high}. Separately a number of authors have considered changes in second order structure of moderate dimensional time series models including auto-covariance and cross-covariance \citep{cho2015multiple}, changes in graphical models \citep{doi:10.1080/10618600.2017.1302340,6854087} and changes in network structure \citep{wang2018optimal}.

The problem of detecting changes in the covariance structure has been examined in both the low dimensional and high dimensional setting. In the low dimensional setting ($p<<n$) \cite{chen2004statistical,lavielle2006detection} utilise a likelihood based test statistic and the Schwarz Information Criterion (SIC) to detect changes in covariance of normally distributed data. \cite{aue2009break} consider a nonparameteric test statistic for changes in the covariance of linear and non-linear multivariate time series. \cite{matteson2014nonparametric} study changes in the distribution of (possibly) multivariate time series using a clustering inspired nonparametric test statistic that claims to handle covariances. In the high dimensional setting, \cite{avanesov2018,wang2017optimal} study test statistics based on the distance between sample covariances, utilising the operator norm and $\ell_\infty$ norm respectively.  Crucially all of these approaches are focused on exploring the theoretical aspects of the proposed test statistics rather than the practical implications.

In this work we propose a novel method for detecting changes in the covariance structure of moderate dimensional time series motivated by the practical challenges of implementing the approach for estimating changes in soil. 
In Section \ref{sec::test}, we introduce a test statistic inspired by a distance metric intuitively defined on the space of positive definite matrices. 
The primary advantage of this metric is that under the null hypothesis of no change, it is independent of the underlying covariance structure. 
This is not the case for other methods in the literature which require users to estimate this. 
In Section \ref{sec::rmt}, we study the asymptotic properties of this test statistic when, the dimension of the data is of comparable size to (but still smaller than) the sample size. 
In Section \ref{sec::cpts}, we use these results to propose a new method for detecting multiple changes in the covariance structure of multivariate time series.
In Section \ref{sec::sims}, we study the finite sample performance of the proposed approach on simulated datasets. 
Finally in Section \ref{sec::appl}, we use our method to examine how changes in the covariance structure of pixel intensities can be used to detect changes in surface water.



\section{Two Sample Tests for the Covariance}
\label{sec::test}

Let $X_{1}, \dots, X_{n} \in \mathbb{R}^p$ be independent $p$ dimensional vectors with
\begin{align}
    \text{Cov}\left(X_{i}\right) =  \Sigma_{i,p}, \text{ for } 1 \leq i \leq n. 
\end{align}
where each $\Sigma_{i,p} \in \mathbb{R}^{p \times p}$ is full rank.
Furthermore, let $X_{n,p}$ denote an $n \times p$ matrix defined by 
$X_{n,p} := (X_{1}^T, \dots, X_{n}^T)$.
Our primary interest in this paper is to develop a testing procedure that can identify a change 
in the covariance structure of the data over time.
For now, let us consider the case of a single changepoint.
We compare a null hypothesis of the data sharing the same covariance versus 
an alternative setting that allows a single change at time $\tau$.
Formally we have 
\begin{align}
    H_0 : & \Sigma_0^* = \Sigma_{1,p} = \dots = \Sigma_{n,p}\\ 
    H_1 : &\Sigma_1^* = \Sigma_{1,p} = \dots = \Sigma_{\tau,p} \neq \Sigma_{\tau + 1,p} = \dots = \Sigma_{n,p} = \Sigma_2^*,
\end{align}
where $\tau$ is unknown.
We are interested in distinguishing between the null and alternative hypothesis,
 and under the alternative locating the changepoint $\tau$,
when the 
dimension of the data $p$, is of comparable size to the length of the data, $n$.
In particular we require that for all pairs $n,p$, the set  
\begin{equation}
    \label{eq::set}
    \mathcal{T}_{n,p}(\ell):= \{ t \in \mathbb{Z}^{+} \text{ such that }  \ell < t < n-\ell \} 
\end{equation}
is non empty, 
where $\ell > p$ is a problem dependent positive constant.
Note $\mathcal{T}_{n,p}(\ell)$ defines the set of possible candidate changepoints,
while $\ell$ is the minimum distance between changepoints or minimum segment length. 
Then for each candidate changepoint $t \in \mathcal{T}_{n,p}(\ell)$, 
a two sample test statistic $T(t)$ can be used to determine if the data to the left 
and right of the changepoint have different distributions.
If the two sample test statistic for a candidate exceeds some threshold, 
then we say a change has occured
and  an estimator for $\tau$ is given by the value $t \in \mathcal{T}_{n,p}(\ell)$ that maximises $T(t)$.

Let $\bar{\Sigma}(p,q)$ (or $\bar{\Omega}(p,q)$) be a plug in estimator for the  covariance (or precision) of data 
$X_{(p+1):q}$.
Then to test for a changepoint at time $\tau$, 
we can measure the magnitude of the matrix 
$
    \psi_{\tau,1}\bar{\Sigma}(0, \tau)-\psi_{\tau,2}\bar{\Sigma}(\tau, n),
$
where 
$\{\psi_{\tau,1}\}_{\tau=\ell+1}^{n-\ell}$ and  
$\{\psi_{\tau,2}\}_{\tau=\ell+1}^{n-\ell}$ are sequences of normalizing constants.
If this matrix is large, then there is evidence for a change and vice versa.
This approach is well represented in the literature, 
for instance \cite{wang2017optimal,aue2009break, galeano2007covariance} measure the difference between sample covariance estimates, 
while in the high dimensional setting,
\cite{avanesov2018} measures the difference between debiased graphical LASSO estimates.
Although this approach is intuitive, 
it can be challenging to use in practice.
A good estimator $\bar{\Sigma}(p,q)$ will depend on the true covariance of $X_{p+1:q}$
which implies that the difference matrix above is dependent on the true covariance of $X_{1:n}$. 
As a result, any test statistic based on a difference matrix must be a function of the underlying covariance, $\Sigma_0$, 
and should be corrected to account for this.
For example, \cite{aue2009break,galeano2007covariance} normalize their test statistic using the sample covariance
for the whole data, 
\cite{avanesov2018} use a bootstrap procedure which assumes knowledge of the measure of $X_i$ and 
\cite{wang2017optimal} use a threshold which is a function of $\Sigma_0^*$.
All these approaches require estimating $\Sigma_0$ in practice.
This is impractical under the alternative setting, since estimating the segment covariances 
requires knowledge of the changepoint.

Therefore, it is natural to ask whether there are alternative ways of measuring the distance between covariance metrics.
In the univariate setting, a common approach is to evaluate the logarithm of the ratio of the segment variances 
(\cite{chen1997testing,inclan1994use,killick2010detection}). 
This is in contrast with the change in expectation problem where it is more common to measure the  difference between sample means.
In the variance setting a ratio is more appropriate for two reasons.
Firstly,  since variances are strictly positive, if the underlying variance is quite small then 
the absolute difference between the values will also be small whereas the ratio 
is not affected by the scale.
Secondly, under the null hypothesis of no change, the variances will cancel and the test
statistic will be independent of the variance.
Thus, there is no need to estimate the variance when calculating the threshold.

We propose to extend this ratio idea from the univariate setting to the multivariate setting by studying 
the multivariate ratio matrix,
$R(A, B) := B^{-1}A$,
where $A, B \in \mathbb{R}^{p \times p}$.
Ratio matrices are widely used in multivariate analysis to compare covariance matrices
(\cite{finn1974general}).
In particular, we are often interested in functions of the eigenvalues of these matrices
(\cite{10.2307/2331979,10.2307/2334137,10.2307/2332232}).
Here we are interested in the following test statistic,
\begin{align}
    T(A,B) 
    = \sum_{j=1}^p \left(1 - \lambda_j(R(A,B))\right)^2 
    + \left(1 - \lambda_j^{-1}(R(A,B))\right)^2,
    \label{eq::test}
\end{align}
where $\lambda_j(R(A,B))$ is the $j$th largest eigenvalue of the matrix $R(A,B)$.
The function $T$ has valuable properties that may not be immediately obvious.
\begin{proposition}\label{thm::invariance}
    Let $\Sigma_1, \Sigma_2 \in \mathbb{R}^{p \times p}$,  $Z_1  \in \mathbb{R}^{n_1 \times p}$ and $Z_2 \in \mathbb{R}^{n_2 \times p}$, and define $T$ as in equation \eqref{eq::test},
    then we have that:
    \begin{enumerate}
        \item $T$ is symmetric i.e. $T(\Sigma_1, \Sigma_2) = T(\Sigma_2, \Sigma_1)$; 
        \item $T$ is symmetric with respect to inversion of matrices i.e. 
            $T(\Sigma_{1}, \Sigma_{2}) = T(\Sigma^{-1}_{1}, \Sigma^{-1}_{2})$ ;
        \item 
            If $\Sigma_1 = \Sigma_{0} Z_1^T Z_1 \Sigma_{0} $ and $\Sigma_2 = \Sigma_{0} Z_2^T Z_2 \Sigma_0$, then 
            $T(\Sigma_1, \Sigma_2) = T(Z_1^T Z_1, Z_2^T Z_2)$.
    \end{enumerate}
   \end{proposition}
The symmetry property is important for a changepoint analysis as the segmentation should be the same regardless of whether the data is read forwards or backwards. 
The second property states that $T$ is the same whether we examine the covariance matrix or the precision matrix.
This ensures that differences between both small and large eigenvalues can be detected.
The third property is particularly important as it implies that $T$ provides a test statistic which is independent of the underlying covariance of the data.
In particular, let $\bar{\Sigma}(p,q)$ be the sample covariance estimate for data $X_{p+1:q}$ i.e. 
$$\bar{\Sigma}(p,q) := \frac{1}{q-p}\sum_{i=p+1}^q X_i X_i^T.$$
Under the null hypothesis for any $1\leq p < q \leq n$, 
$X_{p:q} = \Sigma_0^{1/2} Z_{p:q}$ where the covariance of $Z_i$ is the identity matrix.
Then property 3 implies that for any $1 \leq \tau \leq n$, 
$T(X_{1:\tau}^T X_{1:\tau}, X_{\tau+1:n}^TX_{\tau+1:n}) =  
T(Z_{1:\tau}^T Z_{1:\tau}, Z_{\tau+1:n}^TZ_{\tau+1:n}) $
which is independent of $\Sigma_0$.
In other words, under the null hypothesis the underlying covariance cancels out (as occurs with the ratio approach in the univariate variance setting).
Furthermore, due to the square term, $T$ is a positive definite function. 
This is necessary to prevent changes cancelling out. 
These properties are clearly not unique to our chosen test statistic $T$ and in fact,
there are other possible choices (such as $\log^2{x}$).
However we argue that for an alternative function $T'$ to be appropriate in the changepoint setting,
it would also require these properties.
Furthermore, this choice of $T$ allows us to analytically derive relevant quantities such as the limiting moments.

It is both possible and interesting to study the properties of this test statistic in the finite dimensional setting (i.e., where $p$ is fixed).
However in this work, our focus is on problems where the dimension of the data is of comparable size to the length of the data. 
Under this asymptotic setting, the eigenvalues of random matrices (and by extension the properties of $T$) have different limiting behaviour and a proper test should take this into account.
For example, the two sample Likelihood Ratio Test (as used in \cite{galeano2007covariance}) 
 is a function of the log of the determinant of the covariance,
or equivalently the sum of the log eigenvalues.
In the moderate dimensional setting, 
this test has been shown to breakdown due to the differing limiting behaviour \citep{bai2009corrections}.
Therefore 
in the next section, we consider the properties of $T$ as a two sample test, 
and  derive the asymptotic distribution under moderate dimensional asymptotics.
%

\section{Random Matrix Theory}
\label{sec::rmt}
We now describe some foundational concepts in Random Matrix Theory (RMT),
before discussing how these ideas are utilised to idenfity the asymptotic distribution of our test statistic under the null hypothesis.
RMT concerns the study of matrices where each entry is a random variable.
In particular, RMT is often concerned with the behaviour of the eigenvalues and eigenvectors 
of such matrices. 
Interested readers should see \cite{tao2012topics} for an introduction and 
\cite{anderson2010introduction} for a more thorough review.

A key object of study in the field is the Empirical Spectral Distribution (ESD), defined  for 
a $p \times p$ matrix $A$ as  
\begin{align}\label{eq::ESD}
    F^{A}(x) := \frac{1}{p}\sum_{j=1}^p I(\lambda_{p-j}(A) \leq x) 
\end{align}
where $I$ is an indicator function.
In other words the ESD of $A$ is a discrete uniform distribution placed on the eigenvalues of $A$. 
Several authors have established results on the limiting behaviour of the ESD as the dimension 
tends to infinity, the so-called Limiting Spectral Distribution (LSD).
For example, \cite{doi:10.1137/1009001} demonstrate that if the upper triangular entries of
a Hermitian matrix $A$ have mean zero and unit variance, 
then $F^{1/\sqrt{p}A}(x)$ converges to the Wigner semicircular distribution.

The LSD of the ratio matrix $R$, was shown to exist in \cite{yin1983limiting} 
and computed analytically in \cite{silverstein1985limiting}. 
The following two assumptions are sufficient for the LSD of an F matrix to exist.
\begin{assumption}
      \label{assump1}
    Let $X_{n_1,p} \in \mathbb{R}^{n_1 \times p}$ and $X_{n_2,p}\in \mathbb{R}^{n_2 \times p}$ be random matrices
    with independent not necessarily identically distributed entries 
    $\{X_{n_1,i,j}, 1 \leq i \leq n_1, 1 \leq j \leq p\}$ and $\{X_{n_2,k,j}, 1 \leq k \leq n_2, 1 \leq j \leq p\}$ with mean 0, variance 1 and fourth moment $1 + \kappa$.  
    Furthermore, for any fixed $\eta>0$, 
    \begin{align}
        \frac{1}{n_1 p} \sum_{j=1}^p \sum_{i=1}^{n_1} 
        \mathbb{E}|X_{n_1,i,j}|^4\bm{I}(|X_{n_1, j,k}| \geq \eta\sqrt{n_1}) \to 0 \\ 
        \frac{1}{n_2 p} \sum_{j=1}^p \sum_{i=1}^{n_2} 
        \mathbb{E}|X_{n_2,i,j}|^4\bm{I}(|X_{n_2, j,k}| \geq \eta\sqrt{n_2}) \to 0
    \end{align}
    as $n_1, n_2, p$ tend to infinity subject to Assumption \ref{assump2}.
\end{assumption}
\begin{assumption}
      \label{assump2}
The sample sizes $n_1, n_2$, and the dimension $p$ grow to infinity such that 
    $$\gamma_{n_1} := \frac{p}{n_1} \to \gamma_1 \in (0,1), \hspace{1mm} 
      \gamma_{n_2} := \frac{p}{n_2} \to \gamma_2 \in (0,1) \text{ and } 
      \bm{\gamma} := (\gamma_1, \gamma_2).$$
\end{assumption}
We refer to the limiting scheme described in Assumption \ref{assump2} 
as $\bm{n} \to \infty$.

Let $X_{n_1, p}, X_{n_2,p}$ be matrices satisfying Assumptions \ref{assump1} and 
\ref{assump2}. 
These assumptions place restrictions on the mean of the data, 
and the tails of the data.
The mean assumption is standard in the literature. 
If the data has non zero mean, the data should be standardized, 
typically by removing the sample mean (assuming the mean is constant).
The impact of this on our proposed method is examined in Appendix \ref{app:mean_center}.
Note that although the matrices $X_{n_1, p}, X_{n_2,p}$ have identity covariance, 
these results also hold for data with general covariance $\Sigma_p$, 
since by property 3 of Proposition \ref{thm::invariance} the covariance term cancels out  
under the null hypothesis and we do not require knowledge of $\Sigma_p$.
Furthermore, let $F_{\bm{n}}$ denote the ESD of 
$R(\frac{1}{n_1}X_{n_1,p}^TX_{n_1,p}, \frac{1}{n_2}X_{n_2,p}^TX_{n_2,p})$.
Then \cite{silverstein1985limiting} demonstrate that  
$F_{\bm{n}}$ converges almost surely to the non random distribution function
\begin{align}
    F_{\bm{\gamma}} (dx) &= \frac{1-\gamma_2}{2\pi x(\gamma_1 + \gamma_2 x)}\sqrt{(b-x)(x-a)}
I_{[a,b]}(x) dx \text{ as } \bm{n} \to \infty \\
    \text{where }
    h &= \sqrt{\gamma_1 + \gamma_2 - \gamma_1 \gamma_2}, \hspace{4mm}
 a = \frac{(1-h)^2}{(1-\gamma_2)^2},\hspace{4mm}
 b = \frac{(1+h)^2}{(1-\gamma_2)^2}.
\end{align}

\begin{figure}
    \includegraphics[]{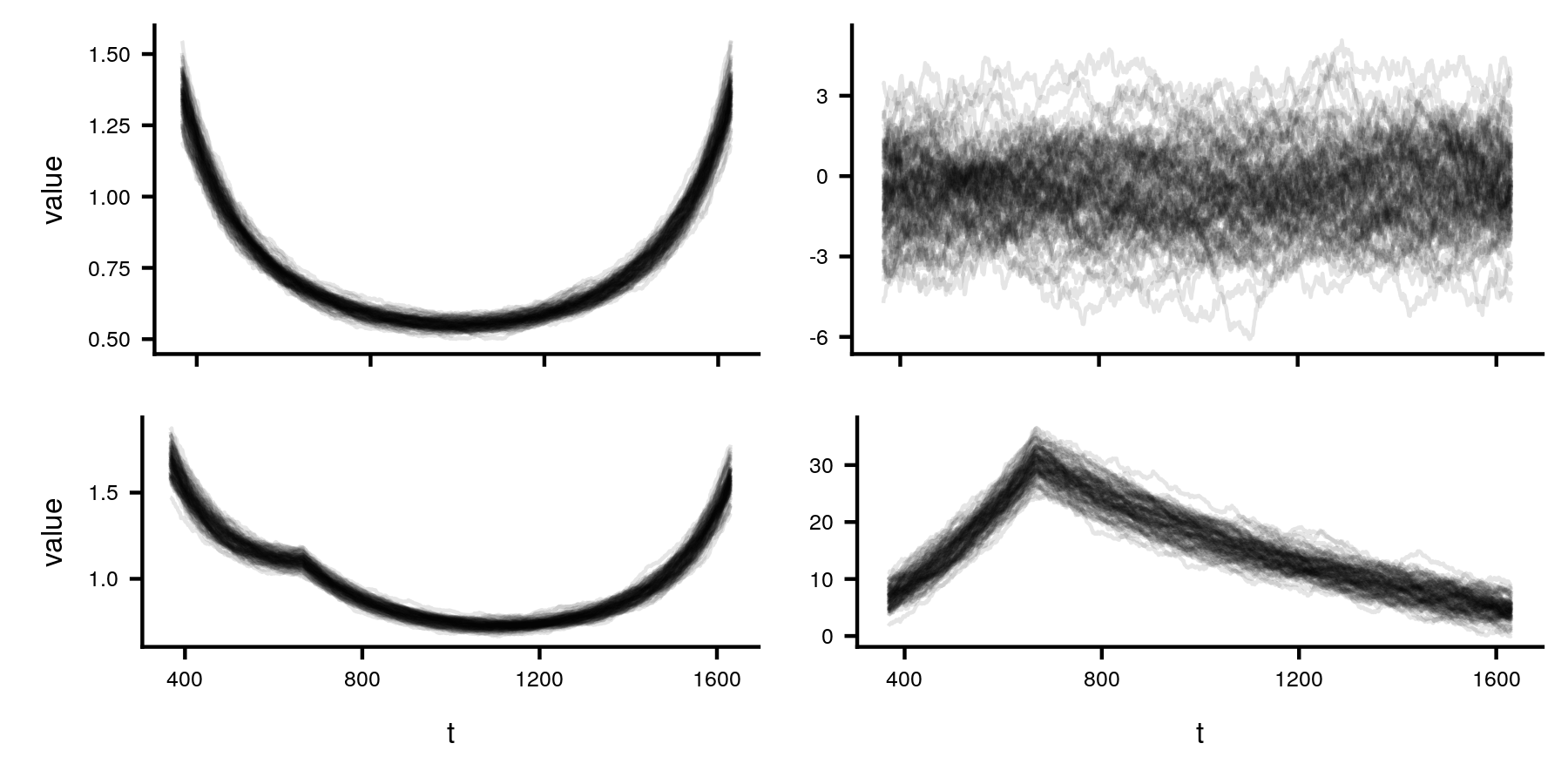}
    \caption[Proposed test statistic before and after standardisation.]{Test statistic $T$ defined in \eqref{eq::test} applied to a 100 different data sets before (left) and after standardisation (right) using \eqref{eq::standard} under the null setting (top) and alternative setting (bottom) with $n=2000$, $p=100$ and $\tau =666$.}
    \label{fig::raw_vs_corrected}
\end{figure}

The LSD, $F_{\bm{\gamma}}$ provides an asymptotic centering term for
functions of the eigenvalues of random ratio matrices. 
In particular, for any function $f$,  we have that,
$$ \mathbb{E}_{F_n}\left(f\right) = \frac{1}{p}\sum_{i=1}^p f\left(\lambda_i\left(R\left(\frac{1}{n_1}X_{n_1,p}^TX_{n_1,p}, \frac{1}{n_2}X_{n_2,p}^TX_{n_2,p}\right)\right)\right)  
\to \int f(x) d F_{\bm{\gamma}}(x) = \mathbb{E}_{F_{\bm{\gamma}}}(f)$$
as $\bm{n} \to \infty$,
by the definition of weak convergence. 
This allows us to account for bias in the statistic as seen in Figure \ref{fig::raw_vs_corrected}.

The rate of convergence of  $\abs{\mathbb{E}_{F_n}\left(f\right) - \mathbb{E}_{F_{\bm{\gamma}}}(f)}$ to zero was studied 
in \cite{zheng2012} and found to be $1/p$.
In particular, the authors establish a central limit theorem for the quantity, 
$$ G_{\bm{n}} (x) := p \left[ F_{\bm{n}}(x) - F_{\bm{\gamma}}(x)\right].$$
We can apply this result to our problem in order to demonstrate that our two sample test statistic converges to a normal distribution with known mean and variance terms.

\begin{theorem}
    \label{thm::bias_var}
    Let $X_{n_1} \in \mathbb{R}^{n_1 \times p}$ and $X_{n_2} \in \mathbb{R}^{n_2 \times p}$ be random matrices
    satisfying Assumptions \ref{assump1} and \ref{assump2} and $T$ be defined as in \eqref{eq::test}. 
    Then we have that as $\bm{n} \to \infty$, 
    \begin{align*}
      T\left(\frac{1}{n_1}X_{n_1,p}^TX_{n_1,p}, \frac{1}{n_2}X_{n_2,p}^TX_{n_2,p}\right) - p\int f^{*}(x) dF_{\bm{\gamma}}(x) \to N(\mu(\bm{\gamma}), \sigma^2(\bm{\gamma}))
    \end{align*}
 where $f^{*}(x) = (1-x)^2 + (1 - 1/x)^2$,
            $\mu(\bm{\gamma}) = 2K_{3,1}\left(1 -  \frac{y_2^2}{h^2}\right) + \frac{2K_{2,1}y_2}{h} +
                2K_{3,2}\left(1 -  \frac{y_1^2}{h^2}\right) + \frac{2K_{2,2}y_1}{h}$, 
    $\sigma^2( \bm{\gamma}) = 
        2(K_{2,1}^2 + 2K_{3,1}^2 + K_{2,2}^2 + 2K_{3,2}^2 +  
        \frac{J_1K_{2,1}}{h} +$ 
        $\frac{J_1K_{2,1}}{h(h^2-1)} + 
        \frac{-J_1K_{3,1}(h^2+1)}{h^2} + 
        \frac{-J_1K_{3,1}}{h^2(h^2-1)} + 
        \frac{J_2K_{2,1} 2h}{(h^2-1)^3} + 
        \frac{J_2K_{3,1}}{h^2} + 
        \frac{J_2K_{3,1}(1-3h^2)}{h^2(h^2-1)^3})$,
    $K_{2,1} =  \frac{2h(1+h^2)}{(1-y_2)^4} -\frac{2h}{(1-y_2)^2}$, 
    $K_{2,2} =  \frac{2h(1+h^2)}{(1-y_1)^4} -\frac{2h}{(1-y_1)^2}$, 
    $K_{3,1} = \frac{h^2}{(1-y_2)^4}$,  
    $K_{3,2} = \frac{h^2}{(1-y_1)^4}$,
    $J_1 = - 2(1-y_2)^2, J_2 = (1-y_2)^4$, 
        $h = \sqrt{y_1 + y_2 - y_1y_2}$, 
    $y_1 = \frac{p}{n_1}$, $y_2 = \frac{p}{n_2}$.

\end{theorem}



Using Theorem \ref{thm::bias_var}, we can properly normalise $T$ such that it can be utilized within a changepoint analysis.
In particular, we have that under the null hypothesis
\begin{align*}
    T(\bar{\Sigma}(0, t), \bar{\Sigma}(t, n)) - p \int f^{*}(x) d F_{\bm{\gamma}_{t/n}} \to 
N(\mu(\bm{\gamma}_{t/n}), \sigma^2(\bm{\gamma}_{t/n}))
\end{align*}
weakly as $n,p$ tend to infinity, where $\gamma_{t/n} := (p/t, p/(n-t))$ and 
$f^{*}$ is as defined in Theorem \ref{thm::bias_var}. 
Thus we utilise the normalised test statistic, $\tilde{T}$,  
\begin{align}
        \label{eq::standard}
\tilde{T}(t) := \sigma^{-1/2}(\bm{\gamma}_{t/n})\left(T(\bar{\Sigma}(0, t), \bar{\Sigma}(t, n)) - p \int f^{*}(x) d F_{\bm{\gamma}_{t/n}} 
 - \mu(\bm{\gamma}_{\tau/n})\right),
\end{align} 
which under the null hypothesis converges pointwise to a standard normal random variable.

The asymptotic moments of the test statistic, $T$,
depend on the parameter $\bm{\gamma}_{t/n}$ and, 
as $t$ approaches $p$ (or equivalently $n-p$) the mean and variance of
the test statistic dramatically increase.
In the context of changepoint analysis, this implies that the mean and variance increase at the edges of the data. 
We note that this is a common feature of changepoint test statistics.
We can significantly reduce the impact of this by the above standardisation.
This can be seen empirically in Figure \ref{fig::raw_vs_corrected}.
After standardisation, the test statistics for the series with no change,
do not appear to have any structure.
Similarly, the test statistics for the series with a change show a clear peak at the changepoint.
Importantly we can now easily distinguish the test statistic under the null and alternative hypotheses, and  this normalization does not require knowledge of the underlying covariance structure.


\section{Practical Considerations}
\label{sec::cpts}
Before we can apply our method to real and simulated data, 
we need to address three practical concerns,
namely we must select a threshold for rejecting the null hypothesis, 
determine an appropriate minimum segment length and 
address the issue of multiple changepoints.

\subsection{Threshold for Detecting a Change}
\label{sec::threhold}
Firstly, 
we need to select an appropriate threshold for rejecting the null hypothesis.
We choose to utilise the asymptotic distribution of the test statistic on a pointwise basis,
that is for each $\ell < t < n- \ell$ we say that $\tilde{T}(t) \approx Z_t$,
where $Z_t$ is a standard normal variable.
This do not take into account whether or not 
we are in the limiting regime and as a result, the method may be unreliable if $p$ is small (indeed we observe this pattern in Section \ref{sec::sims}).
We then use a Bonferroni correction \citep{Haynes2013} to control the probability that any $Z_t$ exceeds a threshold $\alpha$. 
In particular, for a given significance level $\alpha$, 
we reject the null hypothesis  
for a single change in data of length $n$ if 
$\tilde{T}(t) > q(1-\alpha/n)$ for some $\ell < t < n - \ell$,
where $q(\alpha)$ is the $\alpha$th quantile of the standard normal distribution
In the case of multiple changepoints, we use $q(1-2\alpha/n(n+1))$ to account for the 
extra hypothesis tests.
We note that a Bonferroni correction is known to be conservative and as a result, 
using this approach may have poor size (again results from the simulation study validate this concern).
Ideally, one would take account of the strong dependence between consecutive test statistics to 
get a better threshold,
but this is challenging given the non-linear nature of the test statistic.
Further work may wish to investigate whether finite sample results which exploit this dependence can be derived.  Alternatively practitioners could use several different thresholds and ascertain the appropriate threshold for a particular application at hand as demonstrated in \cite{lavielle2005}.



\subsection{Minimum Segment Length}\label{sec:minseglen}
The test statistic proposed relies on an appropriate choice for the minimum segment length parameter, 
$\ell$.  Too small and the covariance estimates in the small segments will elicit false detection, too large and the changepoints will not be localized enough to be useful.

\begin{figure}
    \includegraphics[width=.9\linewidth, height=4cm]{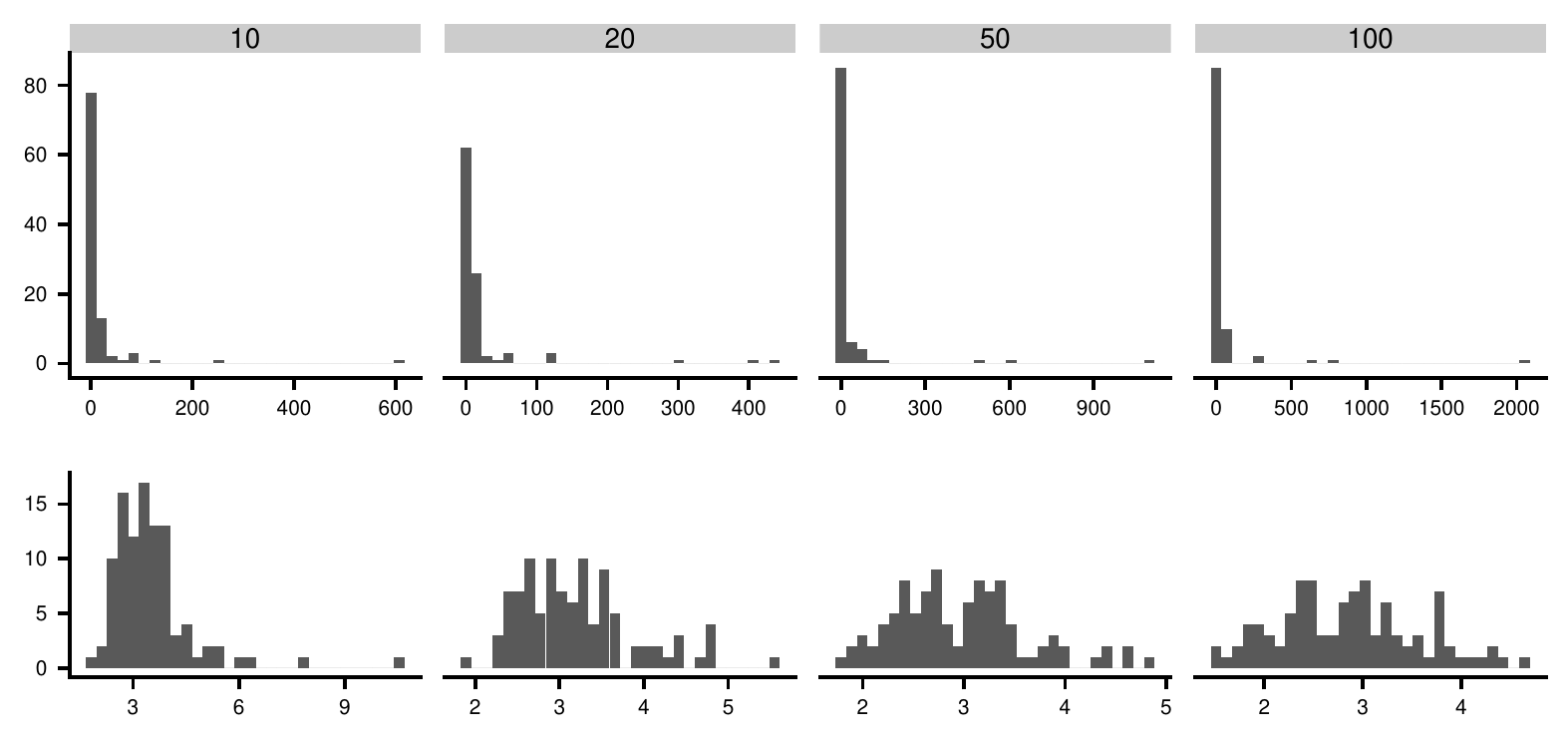}
    \caption[Impact of minimum segment length on distribution of test statistic.]{Histogram of values of $\underset{\ell < t < n - \ell}{\max}\tilde{T}(t)$ 
    applied to 100 datasets of length $n=2000$ with no change for $p=\{10,20,50,100\}$ 
    with $\ell=p$ (top) and $\ell=4p$ (bottom).}
    \label{fig::null_max_hist}
\end{figure}

In many applications, domain specific knowledge may be used to increase this parameter.
However, it is also important to consider smallest value that will give reliable results in the general case. 
The minimum segment length must grow sufficiently fast to ensure that
$\tilde{T}(t)$ converges to a normal distribution. 
Outside the asymptotic regime, it is possible for the ratio matrix to have very large eigenvalues.
Thus for candidate changepoints $t$ close to $p$ (or by symmetry $n-p$), 
the probability of observing spuriously large values of $\tilde{T}(t)$ becomes much larger.
This can be seen in Figure \ref{fig::null_max_hist}.
When $\ell=p$ (the smallest possible value), we observe extremely large values of the test 
statistic, that would make identifying a true change almost impossible.
However when $\ell = 4p$, the test statistic behaves reliably.
Thus we need $p/(p+\ell_{n,p})$ to converge to $\gamma_{\ell} \in (0,1)$ or equivalently
$\ell_{n,p} = \mathcal{O}(p)$ for the asymptotic results to hold.
In Appendix \ref{app:minseglen}, we analyse the effect of different sequences in the finite sample setting via a simulation study.
Based on these results, we recommend  using a default value of $\max\{4p,30\}$, 
however note that for moderate values of $p$, smaller values can be taken without any corresponding decrease in performance.


%

\subsection{Multiple Changepoints}
Finally, we also consider the extension to multiple changes.
In this setting, we have a set of $m$ unknown ordered changepoints, 
$\bm{\tau} := \{0 = \tau_0, \tau_1, \dots \tau_m, \tau_{m+1}=n\}$ 
such that,
   $ X_i \sim \mathcal{N}(0, \Sigma_i) \text{ where }\Sigma_i = \Sigma_k^*, \hspace{1mm} \tau_k < i \leq \tau_{k+1}, \hspace{1mm}
     1 \leq k \leq m+1,$
 and $\Sigma_i$ is the covariance matrix of the $i$th time-vector.
We are interested in estimating the number of changes $m$, and the set of changepoints $\bm{\tau}$.
The classic approach to this problem,
is to extend a method defined for the single changepoint setting to the multiple changepoint setting,
via an appropriate search method such as 
dynamic programming  or binary segmentation.
For this work, we cannot apply the dynamic programming approach \citep{killick2012optimal},
which minimises the within segment variability through a cost function for each segment.
This is because our distance metric is formulated as a two-sample test and cannot be readily expressed as cost function for a single segment.
Therefore we use the classic binary segmentation procedure \citep{scott1974cluster}.
The binary segmentation method extends a single changepoint test as follows.
Firstly, the test is run on the whole data.
If no change is found then the algorithm terminates.
If a changepoint is found, it is added to the list of estimated changepoints and,
the binary segmentation procedure is then run on the data to the left and right of the candidate change.
This process continues until no more changes are found.
Note the threshold, $v$, and the minimum segment length, $\ell$, remain the same.
We note that a number of extensions of the traditional binary segmentation procedure have been
proposed in recent years \citep{fryzlewicz2014wild, olshen2004circular}.
Although we do not use these search methods in our simulations, 
due to additional optional parameters that affect performance,
it is not difficult to incorporate our proposed test statistic into these adaptations of the original binary segmentation approach. 
The full proposed procedure is described in Algorithm \ref{alg::ratiobinseg}.

\begin{algorithm}
        \SetAlgoLined
            \Input{Data matrix $X$, 
                $(s,e)$,
                $\mathcal{C}$,
                minseglen $\ell$, significance level $\alpha$}
                $v = 1 - \frac{\alpha}{n^2}$\;
             \For{$s + \ell \leq \tau \leq e - \ell$}{
                 $\bm{\gamma} := (\frac{p}{\tau}, \frac{p}{n-\tau})$\;
                 $ \tilde{T}(\tau) := \sigma^{-1/2}(\bm{\gamma})
                 \left(T(\bar{\Sigma}(s,\tau), \bar{\Sigma}(\tau, e)) - p \int f^{*}(x) d F_{\bm{\gamma}} 
 - \mu(\bm{\gamma})\right)$\;
             }
             $\hat{\tau} := \underset{s+\ell<\tau<e-\ell}{\arg \max}\tilde{T}(\tau)$\;
             \If{$ \tilde{T}(\hat{\tau}) > v$}{
                 $\mathcal{C}_l :=$ RatioBinSeg$(X, (s, \tau), \mathcal{C}, \ell, \alpha)$\;
                 $\mathcal{C}_r :=$ RatioBinSeg$(X, (\tau, e), \mathcal{C}, \ell, \alpha)$\;
                $\mathcal{C} = \mathcal{C}\cup\{\hat{\tau}\}\cup\mathcal{C}_l\cup \mathcal{C}_r$\;
            }
              \Output{Set of changepoints $\mathcal{C}$.}
              \caption{Ratio Binary Segmentation(RatioBinSeg)}
              \label{alg::ratiobinseg}
\end{algorithm}

%

\section{Simulations}
\label{sec::sims}
In this section, we compare our method with existing methods in the literature, 
namely the methods of \cite{wang2017optimal,aue2009break, galeano2007covariance}, 
which we refer to as the Aue, Galeano and Wang methods respectively.
We do not consider \cite{avanesov2018} as this method is intended for the high dimensional setting and so would be an unfair comparison.
Software implementing these methods is not currently available and as a result,
we have implemented each of these methods according to the  descriptions in their respective papers.
All methods, simulations, visuals and analysis have been implemented in the R programming language \citep{Rref}.
The code to repeat our experiments is available at 
\url{https://github.com/s-ryan1/Covariance_RMT_simulations}.

Simulation studies in the current literature for changes in covariance structure are very limited.
\cite{wang2017optimal} do not include any simulations.  
\cite{aue2009break,avanesov2018} only consider the single changepoint setting,  
and do not consider random parameters for the changes.
Furthermore to our knowledge, no papers compare the performance of different methods. 
While theoretical results are clearly important, 
it is also necessary to consider the finite sample performance of any estimator,
and we now study the finite sample properties of our approach on simulated datasets.
Further details on the general setup of our simulations are given in Appendix \ref{app:simsetup}.  
Note that the significance thresholds for each method are set to be favourable to competing methods and 
we anticipate that performance would decrease in real data.

We begin by analysing the performance of our approach (which we refer to as the Ratio method) in 
the single changepoint setting. 
This allows us to directly examine the finite sample properties of the method, 
such as the power and size, as well as
investigate how violations to our assumptions, such as autocorrelation and heavy tailed errors impact the method.
We then compare our approach with current state of the art methods for 
detecting multiple changepoints.
Results for assessing the chosen default values for the minimum segment length parameter in Section \ref{sec::cpts}, 
as well as a comparison of different methods in the single changepoint setting,
are given in Appendix \ref{app:moresims}.  
These demonstrate that the Ratio and Aue methods  are well peaked whereas the Wang method is not leading to accurate changepoint localization.  
In both settings, the localization of the changepoints is more accurate for our approach than the \cite{aue2009break} method 
whilst also being applicable to larger values of $p$.
However, we note that for smaller dimensions, $p$, the Aue method would likely be more accurate due to the Bonferroni correction over correcting.
Finally, comparisons for the Ratio method based on whether or not the mean is known (under the null and alternate hypotheses) are provided in Appendix \ref{app:moresims}.
These results show that centering the data by subtracting the sample mean has a small impact on the performance of the method.

\paragraph{Performance Metrics}
In the single changepoint setting, we are interested in whether the Ratio approach provides a valid hypothesis test.
Therefore, for a given set of simulations, we measure how often the method incorrectly rejects the null (Type 1 error) and 
how often the method fails to correctly reject the null (Type 2 error).
Furthermore, under the alternative hypothesis, we measure the absolute difference between the estimated changepoint and the true change,
and refer to this throughout as the Changepoint Error.
For the multiple changepoint setting, 
we use $\bm{\tau} := \{\tau_1, \dots, \tau_m\}$ and 
$\hat{\bm{\tau}} :=\{\hat{\tau}_1, \dots, \hat{\tau}_{\hat{m}}\}$
 to denote the set of true changepoints and  the  set of estimated changepoints respectively.
We say that the changepoint $\tau_i$ has been detected correctly if  
$  |\hat{\tau}_j - \tau_i| \leq h$ for some $1 \leq j \leq \hat{m}$ and  
denote the set of correctly estimated changes by $\bm{\tau}_c$.
Then we define the true discovery rate (TDR) and  false discovery rate (FDR) as follows,
$$ TDR := \frac{|\bm{\tau}_c|}{|\bm{\tau}|}, \hspace{1mm} 
FDR := \frac{|\hat{\bm{\tau}}| - |\bm{\tau}_c|}{|\hat{\bm{\tau}}|}.$$ 
A perfect method will have a TDR of 1 and FDR of 0. 
We set $h=20$ although it should be noted that in reality the desired 
accuracy would be application specific and dependent on the minimum segment length $l$.  Although, whilst the specific values vary with $h$ the conclusions of the study do not.
We also consider whether or not the resulting segmentation allows us to estimate the true 
underlying covariance matrices, and define the mean absolute error (MAE) as 
$$MAE := \frac{1}{n}\sum_{i=1}^n \norm{\hat{\Sigma}_i - \Sigma_i}_1.$$
\subsection{Finite sample properties}
\begin{figure}
    \includegraphics[width=.45\linewidth]{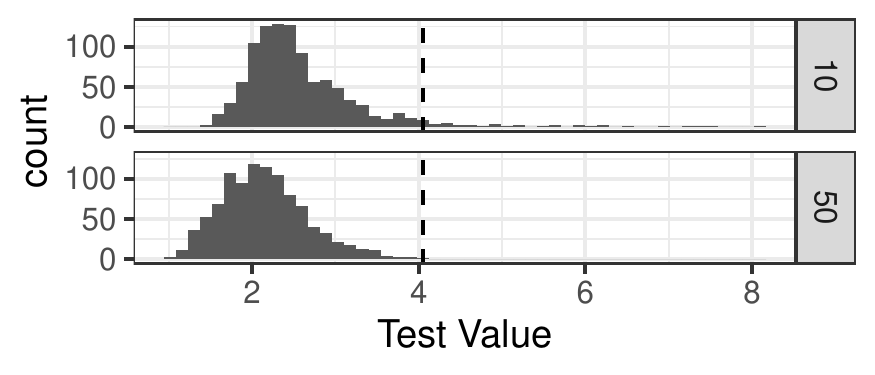}
    \includegraphics[width=.45\linewidth]{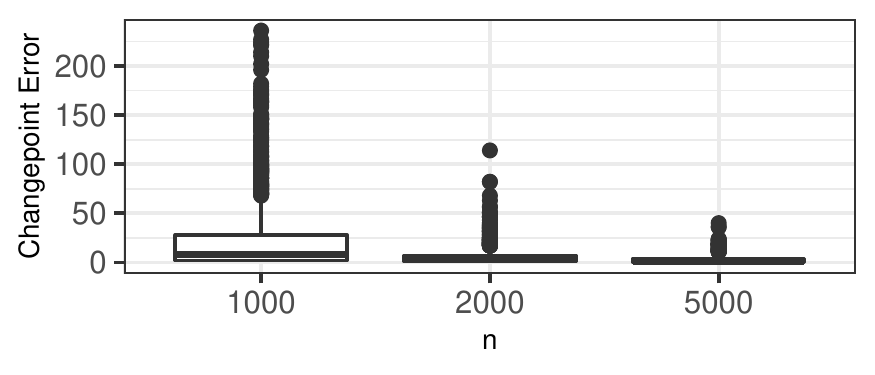}\\
    \includegraphics[width=.45\linewidth]{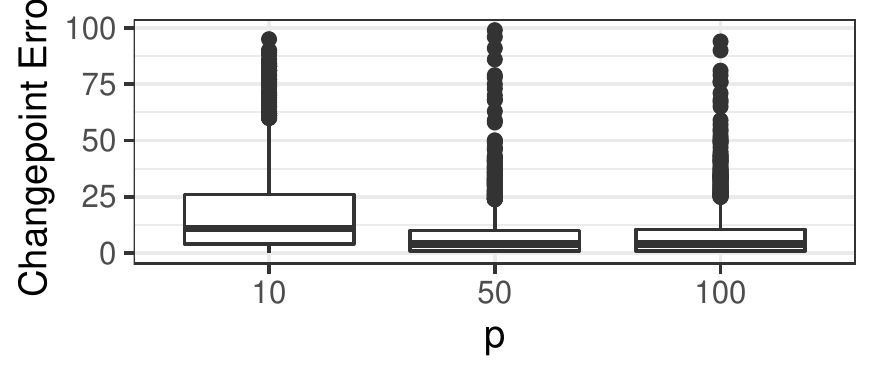}
    \includegraphics[width=.45\linewidth]{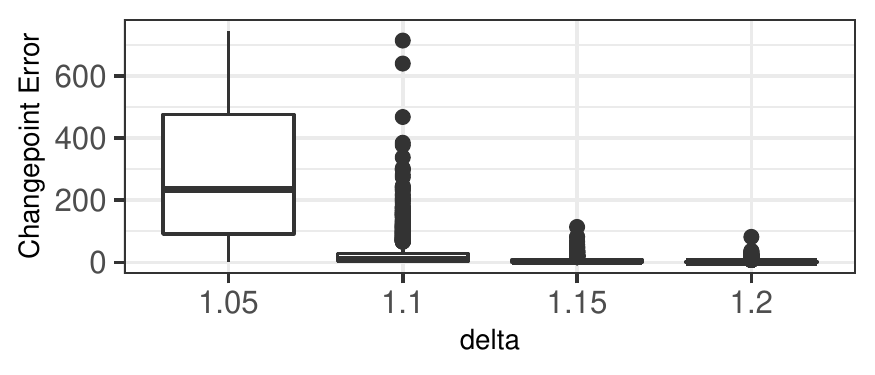}
    \caption{Clockwise from Top Left: Test values for $p=10,50$, dashed line indicates threshold. 
    For higher dimensions we observe less outliers.
    Changepoint error under increasing data length ($n$), dimension ($p$) and size of change ($\delta$).
    The method becomes more accurate as these increase.}
    \label{fig::finite_sample_normal}
\end{figure}

We begin by examining the performance of our proposed approach on normally distributed 
datasets with length $n = \{500,1000,2000,5000\}$ and dimension $p=\{10,50,100\}$.
Based on results from Section \ref{sec:minseglen}, the minimum segment length was set to $4p$.
For each $n,p$ pair, 
we generated 1000 datasets with a single change at time $n/2$ as follows
\begin{align}
    X_i \sim 
    \mathcal{N}(0, I_p) \text{ for } 1 \leq i \leq n/2 \text{ and }
    X_i \sim 
    \delta\mathcal{N}(0,I_p) \text{ for } i > n/2, 
\end{align}
 where $I_p$ is the identity matrix of dimension $p$ and
$ \delta = \{1,1.05, 1.1, 1.15, 1.2\}$.
Note since our approach is invariant to the covariance of the data under the null hypothesis,  
this is equivalent to generating the data with some unknown covariance.
We use the approach described in Section \ref{sec::threhold} to select the threshold 
with $\alpha=.05$.
A histogram of the test statistic values under the null hypothesis $(\delta=1)$ 
for $n=2000$ and $p=10,50$ is shown in Figure \ref{fig::finite_sample_normal}. 
For $p=10$, we observe large test values, 
however this effect is not present for $p=50$,
indicating that we have not entered the limiting regime 
when $p=10$.
We computed the FPR for each $n,p$ pair and 
the results are shown in Table \ref{tab::FPR_TPR}.
Across all dimensions, we observe low numbers of false positives.
In particular, for $p=50$ and $p=100$ the method is conservative.
For $p=10$, we find that the test appears to have good size,
however this difference may be explained by the fact that we have not entered the limiting regime.
We measure the power and accuracy of our method via the True Positive Rate (TPR) 
and the absolute difference between the estimated 
change and the true change (Changepoint Error).
The TPR is given in Table \ref{tab::FPR_TPR} for $\delta=1.1$.
As $n,p$ increases the probability of detection increases.
However for smaller values of $n/p$, the method can have less power, 
such as $n=1000,p=100$ and $n=500,p=50$. 
In these cases, using less data gives a better detection rate,
implying that the method inefficiently utilises the data
and a high dimensional approach may be preferable.
Changepoint errors are given in Figure \ref{fig::finite_sample_normal}.
As $n,p,\delta$ increase, the method  more accurately locates the changepoint,
as we would expect.

\begin{figure}
    \includegraphics[width=.5\linewidth]{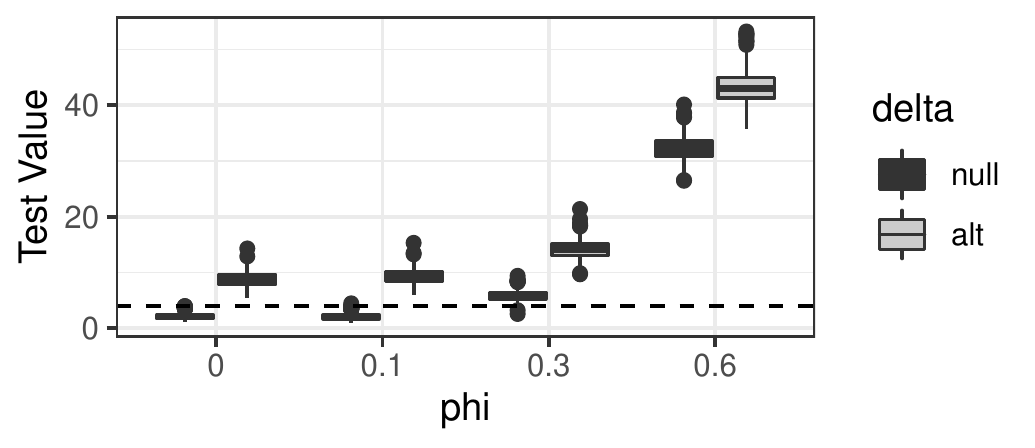}
    \includegraphics[width=.4\linewidth]{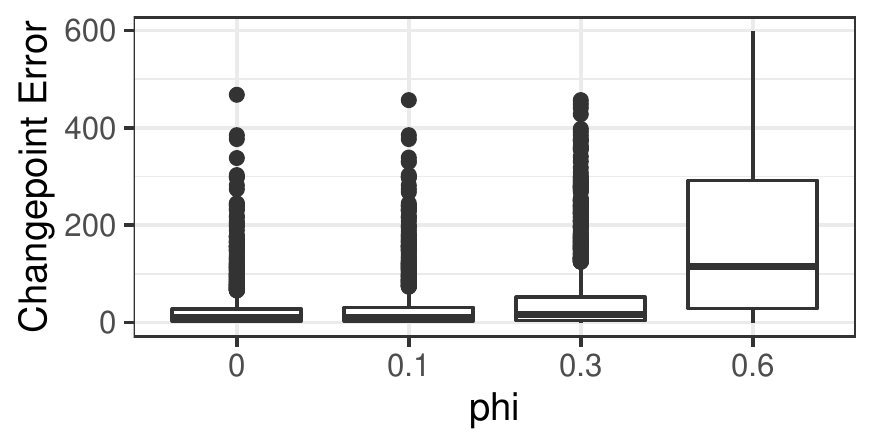}
    \\
    \includegraphics[width=.5\linewidth]{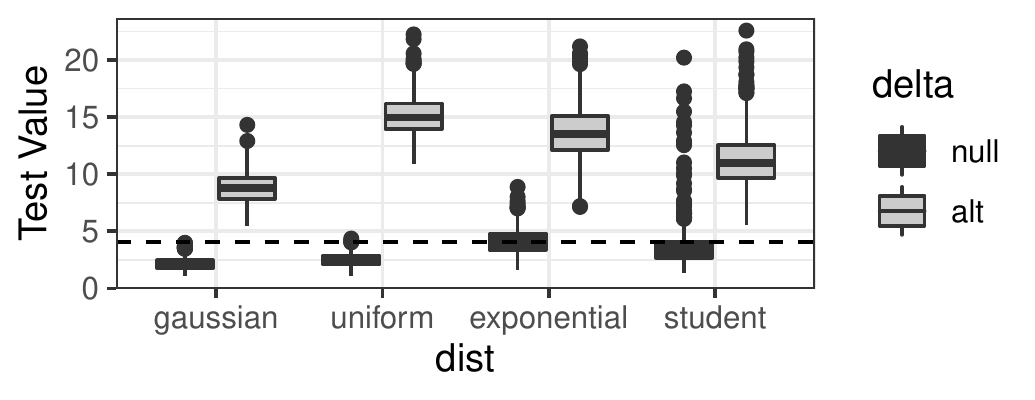}
    \includegraphics[width=.4\linewidth]{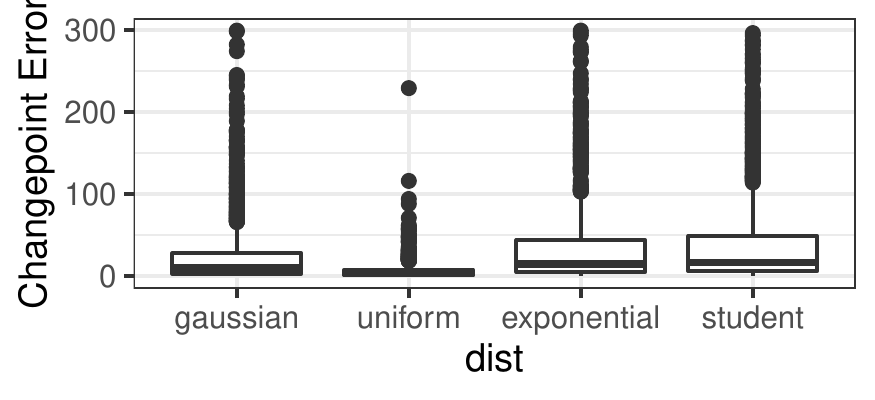}
    \caption{Clockwise from Top Left: 
    Test statistic values for AR(1) data with $\delta=1,1.1$, dashed line indicates threshold. 
    Changepoint Error for AR(1) data ($\delta=1.1)$. 
    Changepoint Error for different error distributions. 
    Test values for different error distributions, dashed line indicates threshold.
    As autocorrelation and probability of outliers increases the method becomes less accurate.
    }
    \label{fig::finite_sample_ts_plus_model}
\end{figure}

\paragraph{Serial Dependence}
Our method does not allow for dependence between succesive data points.
To measure the impact of serial dependence, 
we generated data, 
\begin{align} 
    X_i \sim 
        \phi X_{i-1} + \epsilon_i \text{ for } 1 \leq i \leq n/2 \text{ and }
    X_i \sim 
        \phi X_{i-1} + \delta \epsilon_i \text{ for }i > n/2
\end{align}
where $\epsilon_i \sim \mathcal{N}(0, I_p)$, $n=2000$, $p=50$ and 
$\delta=\{1, 1.1, 1.2\}$ and $\phi = \{0,.1,.3,.6,.9\}$.
Results from this analysis are shown in Figure \ref{fig::finite_sample_ts_plus_model}.
Focusing on the top left plot, we can see that even under the null the test values increase as the autocorrelation increases,
and we find that for $\phi \geq .3$ the proposed threshold is invalid, with FPRs 
of approximately 1.
Thus the test is invalid and will produce spurious false positives.
This is well known in the univariate changepoint literature \citep{Shi2021} and can be mitigated by scaling the threshold by the autocorrelation observed.  

Similarly the power and accuracy of our method decreases when changepoints are present.
Figure \ref{fig::finite_sample_ts_plus_model} shows
 the separation between test statistic value under the null and alternative hypothesis.
If these values are well separated then the method will have good power given a valid threshold.
We find that the separation between the null and alternative distributions decreases and thus changepoint error increases as $\phi$ increases.
These results show that as autocorrelation increases,
our method becomes less accurate as expected.

\paragraph{Model Misspecification}
Our method places assumptions on 
the data which may not hold in practice.
To measure the impact of this we generated data 
\begin{align}
    Y_i =  
\begin{cases}
    X_i - \mu_{f^k} \text{ where } x_{ij} \sim F_k \text{ for } 1 \leq i \leq n/2,  1 \leq j \leq p\\
    \delta(X_i - \mu_{f^k}) \text{ where } x_{ij} \sim F_k \text{ for } i > n/2,  1 \leq j \leq p
\end{cases}
\end{align}
$F_1=\mathcal{N}(0,1), F_2 =$Uniform$(-1/2,1/2)$, $F_3=$Exponential(1), $F_4=$Student t$(5)$, 
$n=2000$, $p=50$ and $\delta = {1,1.1,1.2}$.
Note the the exponential and  t-distributions do not satisfy assumption \ref{assump1} 
and as a result, the threshold is invalid producing FPRs of  .512 and .248 respectively.
Interestingly, although the t-distribution has heavier tailed errors than the exponential distribution,
it gives a lower FPR.
This is likely due to the skewness of the exponential distribution.
We find that the method has less power for the heavier tailed distributions (Figure \ref{fig::finite_sample_ts_plus_model}), 
as again, the distributions under the null and alternative overlap more. 
This pattern is repeated for the changepoint error.  
Thus, the method will be less accurate in the presence of heavy tailed errors.
\begin{table}[ht]
    \scriptsize
    \centering
    \begin{tabular}{rr|rrrr|rrrr|rrrr}
          \hline
          \hline
              &   &     & FDR   &      &     &       & MAE   &     &      & TDR   &    \\
            p & n & Aue & Ratio & Wang & Galeano & Aue & Ratio  & Wang& Galeano& Aue & Ratio & Wang & Galeano\\
             \hline
3       & 500  & 0.38 & \textbf{0.27} & 0.63 & 0.46&\textbf{26.05} &      26.53   & 42.82  &  37.99& \textbf{0.55} &     0.38  & 0.51& 0.20\\
        & 1000 & 0.49 & \textbf{0.29} & 0.77 & 0.51&    18.44  & \textbf{ 15.36}  & 36.81  &  27.94& \textbf{0.58} &     0.48  & 0.48& 0.23 \\
        & 2000 & 0.54 & \textbf{0.31} & 0.86 & 0.54&    12.78  & \textbf{  7.65}  & 32.14  &  21.46& \textbf{0.63} &     0.60  & 0.44& 0.27 \\
        & 5000 & 0.59 & \textbf{0.31} & 0.90 & 0.57&     7.95  & \textbf{  2.93}  & 23.85  &  13.97&     0.65  & \textbf{0.68} & 0.45& 0.31 \\ 
10      & 500  & 0.25 & \textbf{0.16} & 0.28 & 0.46&     304   & \textbf{303.68}  & 316.80 & 541.90& \textbf{0.56} &     0.55  & 0.51& 0.22 \\ 
        & 1000 & 0.32 & \textbf{0.13} & 0.43 & 0.48&    167.59 & \textbf{127.72}  & 243.70 & 413.23&     0.69  & \textbf{0.75} & 0.54& 0.26  \\
        & 2000 & 0.34 & \textbf{0.10} & 0.53 & 0.51&     99.52 & \textbf{ 46.44}  & 186.25 & 292.50&     0.76  & \textbf{0.87} & 0.58& 0.29 \\
        & 5000 & 0.37 & \textbf{0.09} & 0.61 & 0.57&     58.09 & \textbf{ 16.62}  & 123.34 & 189.12&     0.80  & \textbf{0.91} & 0.61& 0.32 \\
30      & 2000 &      & \textbf{0.02} & 0.31 & 0.47&           & \textbf{143.71}  & 1096.3 &1550.72&           & \textbf{0.98} & 0.45& 0.36 \\
        & 5000 &      & \textbf{0.02} & 0.32 & 0.52&           & \textbf{ 51.15}  & 523.25 & 913.45&           & \textbf{0.98} & 0.55& 0.39 \\
100     & 5000 &      & \textbf{0.00} & 0.44 & 0.51&           & \textbf{209.36}  & 7895.1 &5969.70&           & \textbf{1.00} & 0.32& 0.48 \\
            \hline
    \end{tabular}
    \caption{Results from multiple changepoint simulations based on Ratio constraints described in Appendix \ref{app:dgm}. For smaller values of $p$, the Ratio provides lower FPR and lower TPR than the Aue method, indicating a tradeoff between the methods. However for larger values of $p$ the Ratio method is the top performer. Confidence intervals for mean values are provided in Table \ref{tab:se_51}.}
\end{table}

\subsection{Multiple Change Points}
\label{sec::multiple_cpts_sims}
We now compare the Ratio approach with other methods on simulated data sets with multiple changepoints.
We consider datasets with 4 changepoints,
uniformly sampled with minimum segment length $p\log{n}$, 
where $p = \{3,10,30,100\}$ and $n=\{200,500,1000,2000,5000\}$.
Covariance matrices are generated so that the distance between consecutive covariance matrices is sufficiently large to detect a change.
We consider two separate metrics,
$$d_1(A,B) = \sum_{j=1}^p \lambda_j^2(A-B) \text{ and } 
d_2(A,B) = \sum_{j=1}^p (\lambda_j^2(A^{-1}B) - 1)^2, $$
where $d_1$ matches the assumptions from \cite{wang2017optimal}, 
while $d_2$ matches the distance metric used by $T$.
As such, the first set of simulations should favour the Wang method.
Full details for how the covariances were generated are provided in Appendix \ref{app:dgm}.
For each $(n,p)$ pair and distance metric, we generated 1000 datasets and applied our method, the Aue (\cite{aue2009break}) method and the Wang (\cite{wang2017optimal}) method to each dataset. 
Due to its computational complexity, we do not run the Aue method for $p > 10$.
Using the resulting segmentations, we then calculated the error metrics for each method. 
The worst performers across all metrics are the Galeano and Wang methods.
Notably the true positive rate for the Wang method decreases as $p$ grows.
This is in striking contrast with the other methods which become more accurate for larger values 
of $p$ as one may expect.
This may be due to the fact that
the Wang method only considers the first principal component of the difference matrix,
ignoring the remainder of the spectrum or the bias issue identified in Appendix \ref{app:single_change}.
The Galeano performance is equally surprising as given results in the single changepoint case (Appendix \ref{app:single_change})
we would anticipate it to be the best performer. 
Furthermore, these methods also have the highest false positive rate,
indicating that adapting the threshold would not lead to more accurate changepoint estimates.
The Ratio and Aue methods are more closely matched. 
We can see that the Ratio method is the more conservative of the two, 
producing a lower FDR and corresponding lower TDR when $p$ and $n$ are smaller.
This is unsurprising since our results in the single changepoint case found that 
the Ratio method can be less reliable when $p<10$.
For scenarios with $p >30$, the Ratio approach is extremely accurate.
This indicates that for problems with smaller datasets, 
the Aue method may be preferable, while our approach is suitable for larger datasets.
\begin{table}[ht]
    \scriptsize
    \centering
    \begin{tabular}{rr|rrrr|rrrr|rrrr}
        \hline
        \hline
        &   &     & FDR   &      &     &  & MAE   &     &   &   & TDR   &   & \\
        p & n & Aue & Ratio & Wang & Galeano & Aue & Ratio & Wang& Galeano &Aue & Ratio & Wang & Galeano\\
        \hline
        3 & 500 & 0.30 & $\bm{0.10}$ & 0.66 & 0.94 & 21.21       & $\bm{ 16.99}$  & 41.86    &   40.47 &$\bm{0.75}$ &      0.64    & 0.41  & 0.02 \\
        3 & 1000& 0.40 & $\bm{0.14}$ & 0.81 & 0.81 & 16.40       & $\bm{  9.94}$  & 37.33    &   32.33 &$\bm{0.77}$ &      0.72    & 0.35  & 0.05 \\ 
        3 & 2000& 0.47 & $\bm{0.17}$ & 0.88 & 0.68 & 12.14       & $\bm{  5.49}$  & 30.59    &   25.83 &$    0.77$  &      0.77    & 0.3   & 0.09\\ 
        3 & 5000& 0.52 & $\bm{0.20}$ & 0.92 & 0.51 & 7.54        & $\bm{  2.43}$  & 22.21    &   19.51 &0.77        & $\bm{0.81}$  & 0.28  & 0.16\\ 
        10& 500 & 0.36 & $\bm{0.29}$ & 0.57 & 1.00 &$\bm{226.71}$& $244.77$  & 265.47   &  329.02 &$\bm{0.35}$ &      0.28    & 0.19  & 0.00\\ 
        10&1000 & 0.45 & $\bm{0.25}$ & 0.73 & 1.00 &141.09& $\bm{138.85}$  & 201.80   &  253.15 &$\bm{0.46}$ &      0.41    & 0.14  & 0.00\\ 
        10&2000 & 0.48 & $\bm{0.25}$ & 0.80 & 0.99 & 91.04       & $\bm{ 68.39}$  & 147.13   &  206.24 &$0.54$ &      $\bm{0.55}$  & 0.13  & 0.00\\ 
        10&5000 & 0.48 & $\bm{0.20}$ & 0.81 & 0.95 & 50.98       & $\bm{ 21.07}$  & 90.40    &  165.90 &0.63        & $\bm{0.72}$  & 0.14  & 0.01 \\
        30&2000 &      & $\bm{0.03}$ & 0.84 & 0.96 &             & $\bm{131.02}$  & 1127.84  & 1330.00 &            & $\bm{0.96}$  & 0.05  & 0.01 \\ 
        30&5000 &      & $\bm{0.02}$ & 0.84 & 0.87 &             & $\bm{ 42.41}$  & 671.94   &  925.27 &            & $\bm{0.98}$  & 0.06  & 0.04 \\ 
        100&5000&      & $\bm{0.00}$ & 0.97 & 0.64 &             & $\bm{199.11}$  & 7796.28  & 7036.56 &            & $\bm{1.00}$  & 0.01  & 0.11 \\ 
\hline
\end{tabular}
    \caption{Results from multiple changepoint simulations based on assumptions in \cite{wang2018optimal} described in Appendix \ref{app:dgm}. 
    For smaller values of $p$, the Ratio provides lower FPR and lower TPR than the Aue method, indicating a tradeoff between the methods. However for larger values of $p$ the Ratio method is the top performer. Confidence intervals for mean values are provided in Table \ref{tab:se_52}.}
\end{table}

\section{Detecting changes in moisture levels in soil}
\label{sec::appl}
In this section, we investigate whether changes in the covariance structure of soil data correspond with shifts in the amount of moisture on the soil.
There is significant interest in developing new techniques to better understand how water is absorbed and modelling surface water.
This is an important question and is relevant to a variety of industrial applications such as farming, construction and the oil and gas industry \citep{hillel2003introduction}.
A widely used approach is to place probes at different depths and locations in the soil which measure the level of moisture.
To measure across a site more easily, scientists are investigating the use of cameras to capture the soil surface as a surrogate for moisture.



\begin{figure}[htb]
    \includegraphics{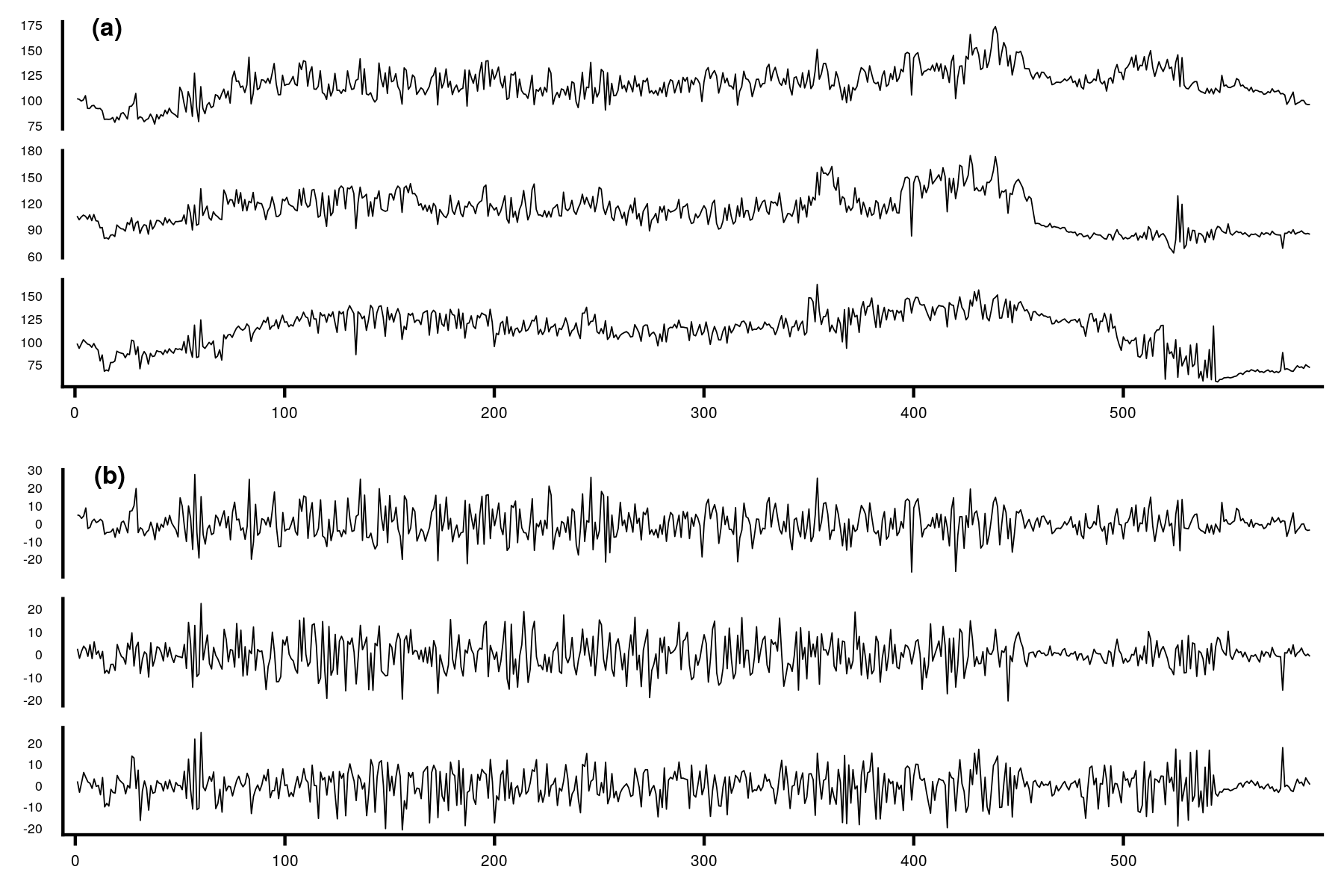}
    \caption[Raw and standardized grayscale intensities for three pixels.]{(a) Raw grayscale intensities for three pixels. (b) Standardised intensities for the same three pixels. }
    \label{fig::soil_data}
\end{figure}

We analyse images from an experiment studying moisture on the surface of the soil.
A camera was placed over a large plot of soil and took a set of 589 pictures over the experiment.
At different times, different amounts of rainfall are simulated and 
the amount of water on the soil surface changes.
We wish to segment the data into periods of dry, wet and surface water.
The intensity of a set of pixels over time is shown in Figure \ref{fig::soil_data} .
We can see that the mean level is clearly nonstationary.
This nonstationary behaviour may be attributed to two causes, 
changes in the background light intensity (due to a cloud passing by) and 
changes in the wetness of the soil which changes how much light is reflected. 
Since changes in the mean intensity are not necessarily associated with changes in the wetness, 
we instead focus on changes in the covariance structure.
If pixels become wet we would expect that the correlation between the pixels should increase as they become more alike as the surface becomes uniformly water instead of the variable soil surface.
Thus changes in the covariance structure of the pixels may correspond with changes in the wetness.

The data consists of 589 images and we analyze two subsets with $p=10,30$ 
The original images are in colour but were transferred to grayscale for computational purposes.
We run a multiple changepoint analysis on the smaller subset using our approach as well as the Aue and Wang methods.
We also ran a multiple changepoint analysis on the larger subset.

In order to analyse the covariance structure of the data, 
we first need to transform the data to have stationary mean.
Estimating the mean of this series is challenging as there is stochastic volatility and 
the smoothness of the function appears to change over time.
As a result, standard smoothing methods such as LOESS and windowed mean estimators may be inappropriate.
We use a Bayesian Trend Filter with Dynamic Shrinkage \citep{kowal2019dynamic}, implemented in the DSP R package \citep{dsp}, which is robust to these issues. 
We then take the residuals.
The transformed data for a subset of the pixels can be seen in Figure \ref{fig::soil_data}.
The minimum segment length is set to 25.
The thresholds for significance for each method were again set to the defaults as discussed in the previous section.
The results of this analysis are shown in Table \ref{tab::cpts}.

To validate our results we worked with scientists  studying this data 
and identified three clear time points where there is a substantial change in the amount of water on the surface at the relevant pixels.
The first change is somewhat gradual going from very dry at time $t=64$ to very wet from time $t=76$.
The second and third changes are more abrupt, with a substantial increase in the amount of water at time $t=350$ and a corresponding sharp decrease at time $t=450$.
The Aue method reports 7 changepoints, 
the Wang method reports 5 changepoints and our method locates 8 changepoints.
All methods detect the first and last changes.
However the Wang method does not detect any change near the second anticipated changepoint.
All of the methods appear to overfit changepoints, in the sense that they report changes that do not correspond with clear changes in the amount of water on the surface.
For our method and the Aue method, 
the majority of these overfitted changes occur when the soil is dry (before t=64 and after t=450).
During these periods the scientists were moving around the site increasing the variability in the amount of light exposure from image to image which may explain these nuisance changes. 

For the larger dataset, the minimum segment length was set to 60 (twice the number of variables) and the thresholds were set to their defaults.
The results were broadly similar for our method and quite different for the Wang method.
Our approach reports 6 changes again detecting the three obvious changes in the video.
We note that the reduced number of changepoints is primarily due to the increased minimum segment length.
The Wang method only reports a single changepoint. 
This drop in reported changes is caused by the largest eigenvalue of the sample covariance being much larger.
As a result, the threshold for detecting a change is 
$3.5$ times larger to account for this and consequently,
it appears that the method loses power.

\begin{table}
    \begin{tabular}{l|l|l}
        Method & Small subset $(p=10)$ & Larger subset$(p=30)$ \\
        \hline 
        Aue & 66, 101, 243, 354, 451, 514, 589  & NA\\ 
        Wang & 52,79, 184, 237, 445 & 445\\
        Ratio & 49, 77, 244, 347, 452, 493, 532,562 & 64, 125, 184, 255, 340, 450, 527
    \end{tabular}
    \caption[Detected changepoints for each of the three methods when applied to the soil image data.]{Detected changepoints for each of the three methods when applied to the soil image data. Note the dimension of the larger subset means the Aue method can detect at most one changepoint.  }
    \label{tab::cpts}
\end{table}

\section{Conclusion}
In this work, we have presented a novel test statistic for detecting changes in the covariance 
structure of moderate dimensional data.
This geometrically inspired test statistic has a number of desirable properties that are not 
features of competitor methods.
Most notably our approach does not require knowledge of the underlying covariance structure.
We utilise results from Random Matrix Theory to derive a limiting distribution for our test statistic. 
The proposed method outperforms other methods on simulated datasets,
in terms of both accuracy in detecting changes and estimation of the underlying covariance 
model.
We then use our method to analyse changes in the amount of surface water on a plot of soil.
We find that our approach is able to detect changes in this dataset that are visible to the eye and 
locates a number of other changes. 
It is not clear whether these changes correspond to true changes in the surface water and
we are investigating this further.

While our method has a number of advantages, it is important to 
recognise some limitations.
Firstly, our method requires calculating the inverse of a matrix at each time point, 
which is a computationally  and memory intensive operation.
As a result, our approach is challenging for larger datasets that can be considered by other
methods, which only require the first principle component.  This could be mitigated by novel solutions to solving inverse matrices alongside GPU computation.
However, as we demonstrate through simulations, there are a wide range of settings where our 
method produces considerable better results for a marginal increase in computational time.
Finally we note that a limitation of our method is that the minimum segment length is bounded below by the dimension of the data.
This means that the method cannot be applied to tall datasets $(p > n)$ or datasets with short segments.

\printbibliography[]

\appendix

\section{Further Details on the Simulation Study}\label{app:simsetup}
\subsection{Random Seed Generation}
To ensure reproducibility, throughout our simulation study we make use of seeds for generating random numbers.
These seeds are chosen to ensure that, where appropriate, 
simulations under different settings are directly comparable.
For example, the data generating mechanisms in Section  5.1 feature a parameter $\delta$ which measures the size of the change.
If the random seed depended on the $\delta$ parameter, 
it would be harder to compare the results across different parameters, 
such as $\delta=1.1$ and $\delta=1.2$, 
since by random chance the noise generated for the former case may be better behaved than for the latter. 
Therefore for all our experiments, 
the random seeds depend on the dimensions of the data, $n$ and $p$, 
however they do not depend on other parameters, 
such as change size or error distribution. 
Finally in Section \ref{sec::multiple_cpts_sims}, the seeds for generating random covariances depends on $p$ but not $n$ 
(since $n$ determines how hard the problem is).

\subsection{Data Generation Mechanism}\label{app:dgm}
We now provide full details for the data generating mechanisms used to simulate data in Section \ref{sec::multiple_cpts_sims}.
In the study, random covariance matrices are generated such that distance between consecutive matrices is sufficiently large.
We incorporate 
constraints on the eigenvalues of the difference and ratio matrix, i.e. 
$$d_1(A,B) = \sum_{j=1}^p \lambda_j^2(A-B) \text{ and } 
d_2(A,B) = \sum_{j=1}^p (\lambda_j^2(A^{-1}B) - 1)^2.$$
The procedure for generating a set of matrices sufficiently far apart with respect to $d_1$ is as follows
\begin{enumerate}
    \item $\Lambda_1 = \text{diag}(\lambda_{1}^1, \dots, \lambda{1}^p)$ where $\lambda_1^j \sim \text{Uniform}(.1,10)$ for $1\leq j \leq p$
    \item For the $k$th segment, generate $\lambda_k^j \sim$ Uniform$(max(\lambda_{k-1}-\kappa_1, .1), \lambda_{k-1}+\kappa_1)$
    \item Then $\lambda_k^J = \lambda_{k-1}^J + \kappa_1$ where $J$ is a random integer uniformly sampled from $\{1, \dots, p\}$. 
    \item Then $\Sigma_k = B^T \Lambda_k B$ where  $\Lambda_k = \text{diag}(\lambda_{k}^1, \dots, \lambda_{k}^p)$
    and $B$ is a random orthonormal matrix.
\end{enumerate}
The procedure for generating a set of matrices sufficiently far apart with respect to $d_2$ is as follows
\begin{enumerate}
    \item $\Lambda_1 = \text{diag}(\lambda_{1}^1, \dots, \lambda{1}^p)$ where $\lambda_1^j \sim \text{Uniform}(.1,10)$ for $1\leq j \leq p$
    \item For the $k$th segment, generate $\lambda_k^j = u^{j+1}-u^{j}$ where $\{u^j\}_{j=1}^p$ is an ordered list of 
        Uniform$(0, Kp)$ random variables. Then $\lambda_k^p = \kappa p - \sum_{j=1}^{p-1} \lambda_{k}^j$.
    \item To ensure that the covariances get both bigger and smaller in scale, we set $\lambda_{k}^j = (1+\lambda_k^j)^{-1^{b^j}}$
        where each $b^j \sim $Bernoulli(1/2)
    \item Then $\lambda_{k}^j = \lambda_k^j \lambda_{k-1}^j$ and $\Lambda_k = \text{diag}(\lambda_{k}^1, \dots, \lambda_{k}^p)$
    \item Finally, $\Sigma_k = B^T \Lambda_k B$ where $B$ is a random orthonormal matrix.
\end{enumerate}


\subsection{Specifications for Competitor Methods}
In Remark 2.1, \cite{aue2009break} state that the asymptotic distribution of their test statistic after 
standardisation can be approximated by a standard normal distribution. 
Therefore we set the threshold for detecting a change to be the $95\%$ quantile or 1.96.
Note that this could be increased, reducing the probability of overfitting changes but also reducing the power of the method.
This approach also requires a plug in estimator for the long run covariance of the vectorized second moment of the data.
For datasets with no temporal structure, 
this long run covariance is exactly the covariance of the vectorized second moment and 
we use the empirical estimate as our plug in estimator.
Note this should improve the performance of the method compared with a generic plug in estimator for the long run covariance.
For examples with temporal dependence, following the recommendation provided in \cite{aue2009break}, 
we use the Bartlett estimator as implemented in \cite{sandwich}.
In both cases, the plug in estimator has dimension $p(p+1)/2$ where $p$ is the dimension of the data, 
and must be invertable implying that $n > p(p+1)/2$.
As a result, we do not include this method in simulations with large datasets.

\cite{wang2017optimal} do not provide a practical default threshold for their method,
instead providing an interval of consistent thresholds which is defined by theoretical 
quantities such as the minimum size of a change, 
the minimum distance between changes and a bound on the tails of the data, $B$. 
A lower bound on the minimum threshold is given by $B^2 \sqrt{p\log{n}}$.
The value $B$ bounds the square root of the largest eigenvalue of the covariance of the underlying data,
which implies the largest eigenvalue is a lower bound for $B$.
Note this value is not available in practice so we approximate this quantity with the largest eigenvalue of the data.
Thus a lower bound for the threshold is given by $ \lambda_{\max}(X) \sqrt{p\log{n}}$.
Again if this value was increased, the method would lose power but be less likely to overfit changes.

\cite{galeano2007covariance} propose a number of methodsm, recommending a Cusum based test statistic in most settings. 
Therefore we compare our approach with the CUSUM method.
The authors demonstrate that their method converges weakly to a standard Brownian Bridge.
We base our threshold on the $95\%$ quantile of the asymptotic distribution. 

\section{Further Simulations}\label{app:moresims}
\subsection{Assessment of minimum segment length}\label{app:minseglen}
\begin{figure}
    \includegraphics[width=.95\linewidth]{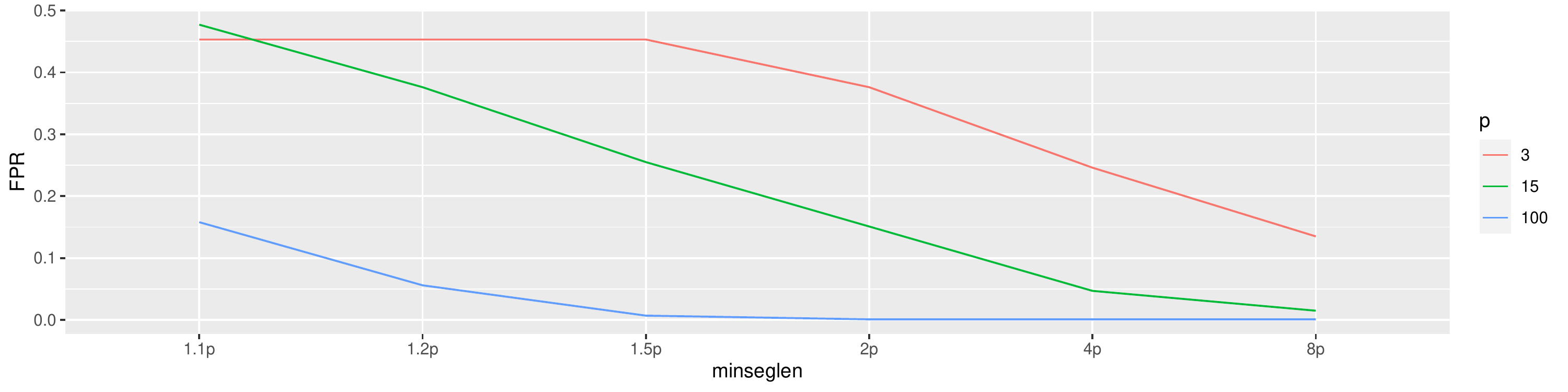}
    \caption[Impact of minimum segment length on false postives rate.]{False Positive Rates for different minimum segment length functions and different values of $p$. We can see that for larger values of $p$, proportionally smaller minimum segment lengths can be used without an increase in FPR.}   \label{fig::FPR} 
\end{figure}

In order to control the false positive rate of the method,
we need appropriate choices of the minimum segment length, $\ell$.
We examined the impact of different functions for $\ell$ on data of different dimensions. 
In particular, we consider datasets with $n=100,p=3$, $n=500,p=15$ and $n=2500,100$.
For each combination, we generated 1000 datasets and applied the Ratio method with 
$\ell = \{1.1p, 1.2p,1.5p,2p,4p,8p\}$.
The results of this analysis can be seen in Figure \ref{fig::FPR}.
As with other simulations the FPR is not well controlled for $p=3$.
For $p=15$, the FPR is well controlled for $\ell \geq 4p$. 
For $p=100$, we can use much less conservative functions for $\ell$, 
with no significant increase in FPR for $\ell=1.5p$.
This is likely due to the fact that when $p=100$, 
we are closer to the limiting regime and the test is better behaved.
As a result of these simulations, 
we recommend setting $\ell=4p$, 
however note that less conservative functions are reasonable for larger values of $p$.

\subsection{Single Changepoint}\label{app:single_change}
\begin{figure}
    \includegraphics[width=.65\textwidth, height=.6\textwidth]{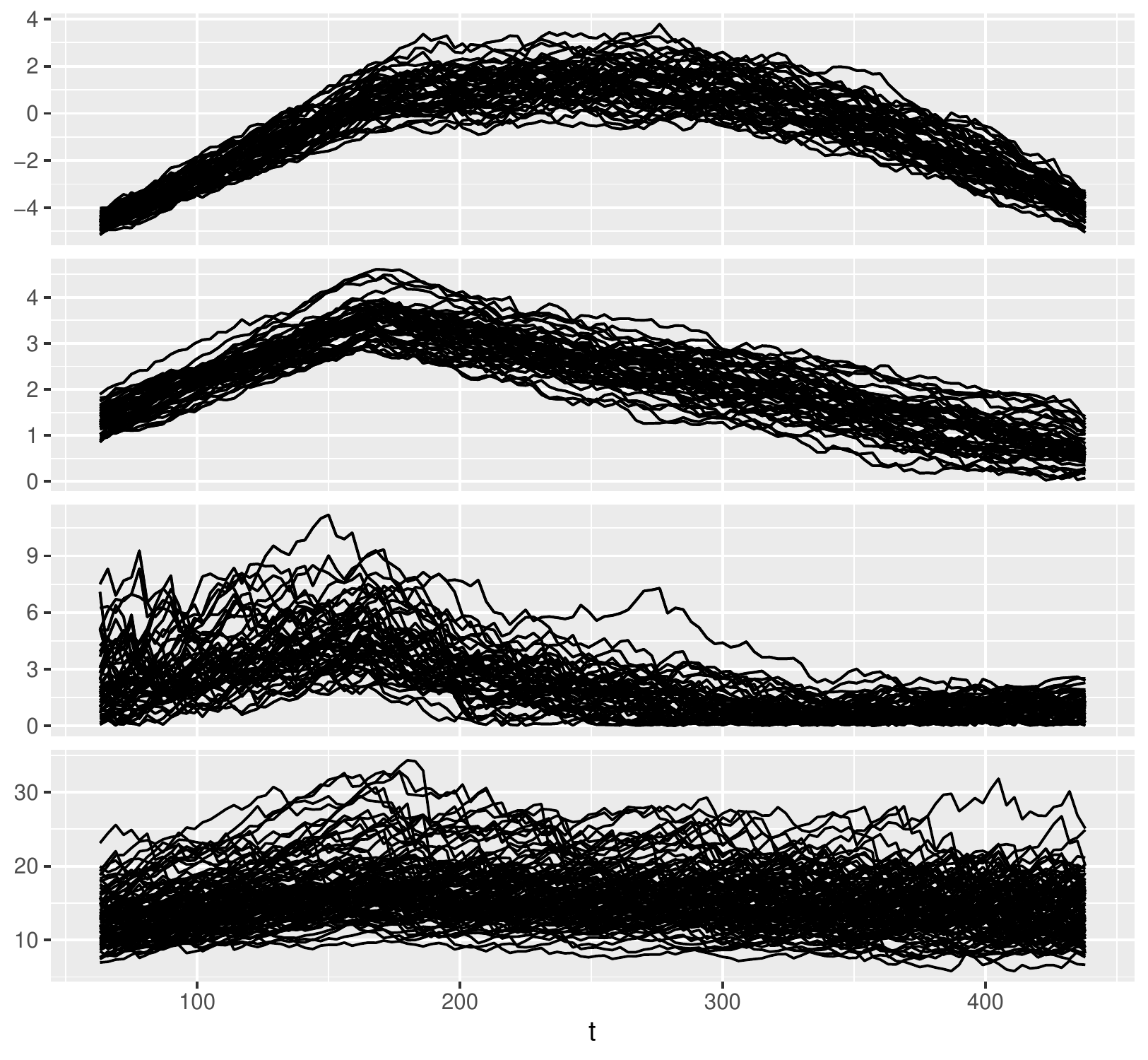}
    \includegraphics[width=.275\textwidth, height=.6\textwidth]{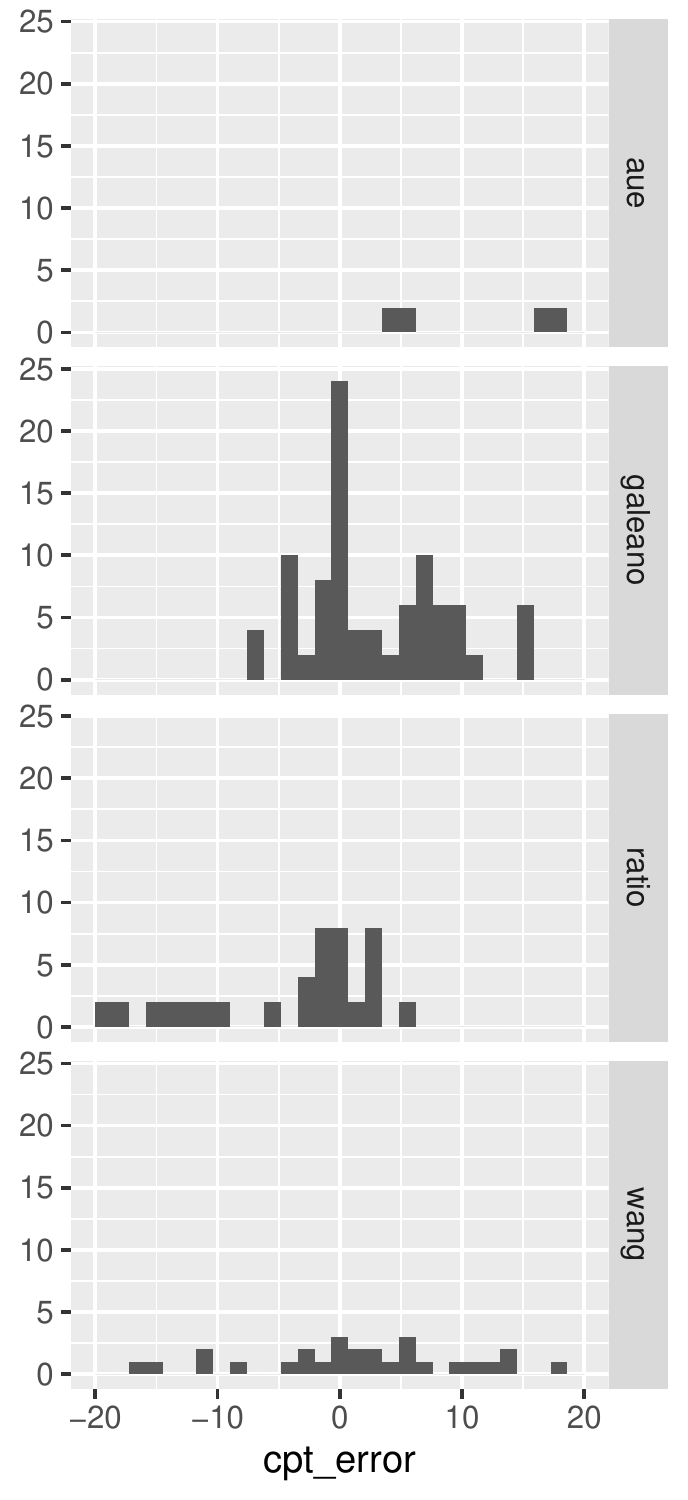}
    \\
    \includegraphics[width=.65\textwidth, height=.6\textwidth]{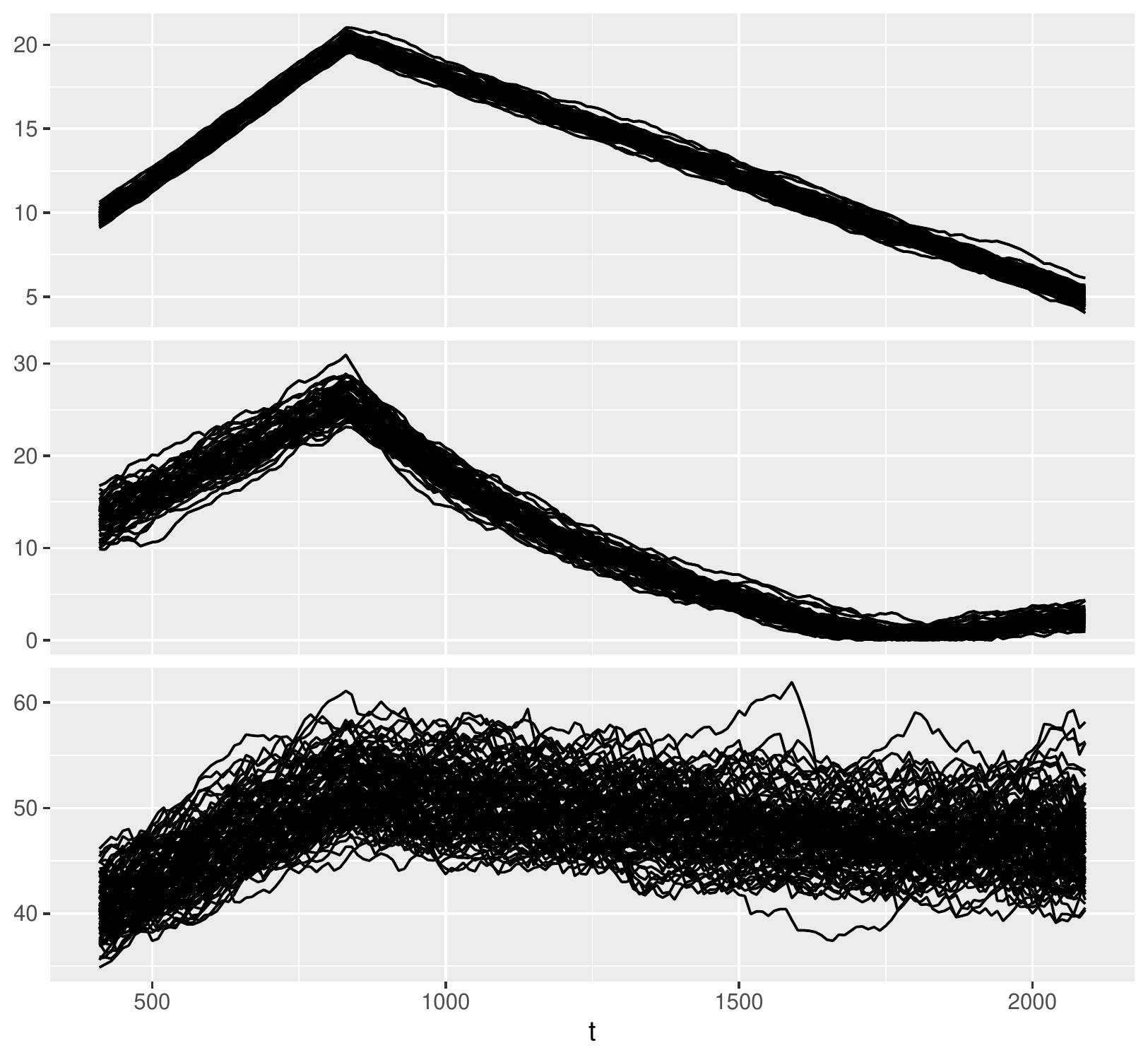}
    \includegraphics[width=.275\textwidth, height=.6\textwidth]{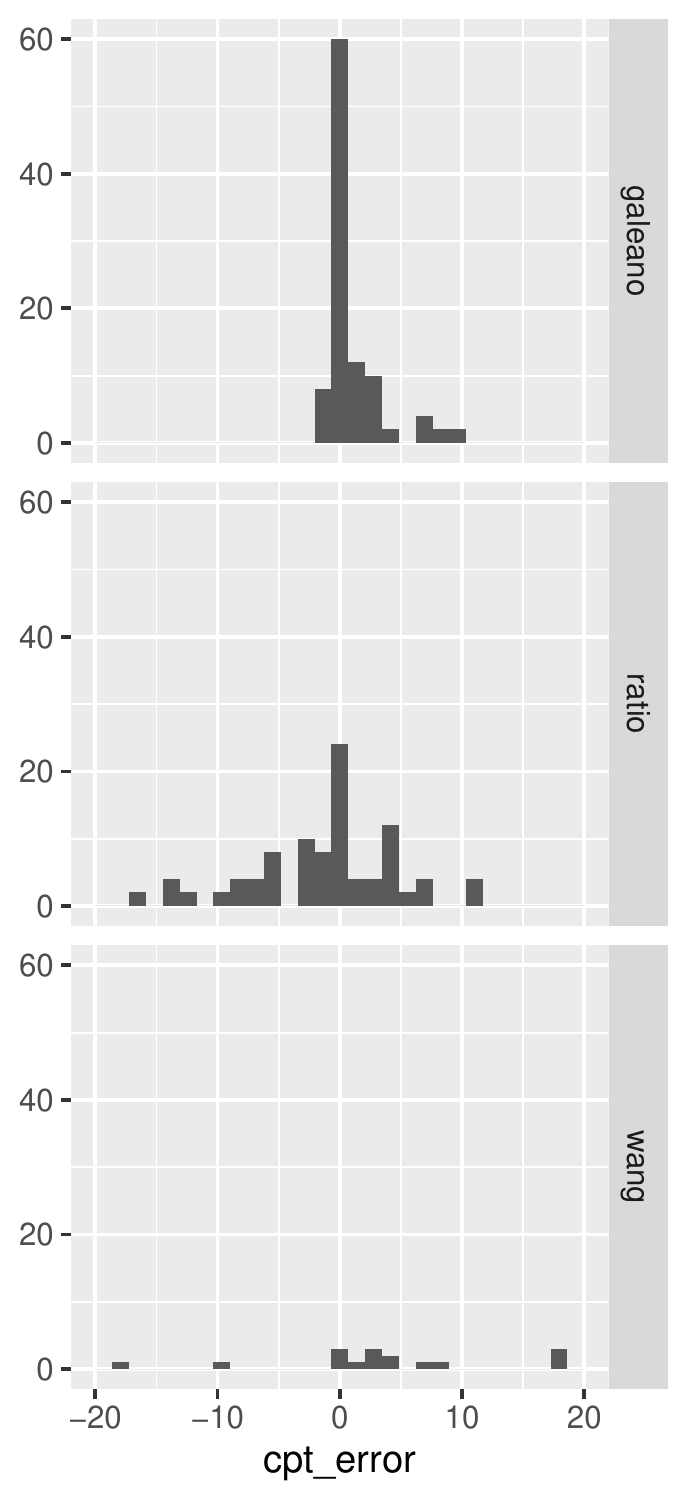}
    \caption[Comparison of different methods for a single changepoint with a fixed location.]{
        (a) Test statistic at each time point from a 100 different data sets under the alternative setting with $p=15$, $n=500$ and a changepoint at $n/3$. (b) Histogram of the difference between the estimated changepoint location and the true changepoint. 
    (c) Same as (a) for $p=100$ and $n=2000$. (d) Same as (b) for $p=100$ and $n=2000$.} 
    \label{fig::fixed_cpt}
\end{figure}

We now compare our approach with the methods mentioned in Section \ref{sec::sims}.
For all our simulated examples, we let the minimum segment length or distance between changes
be $p\log{n}$ as this is required by the wang method.
We compare the four approaches on a set of 100 datasets with a change at $\tau=\lfloor n/3 \rfloor$.
We consider two settings with
the first case having a moderate value for $p$ ($p=15,n=500$), 
and the second case having a larger value for $p$ ($p=100, n=2000$).
Importantly in the second setting we should be closer to the asymptotic regime for our method as 
$n$ and $p$ are larger.
For each dataset we computed the test statistic as well as the difference between the truth and 
the changepoint estimates for each method.
Note that the Aue method is not computable for the $p=100$ case, 
and as a result is not included for this case.
The results of this simulation can be seen in Figure \ref{fig::fixed_cpt}.

In both the small and large $p$ cases, the Galeano  method clearly outperforms the other methods,
providing the lowest changepoint error.
Looking at Figure \ref{fig::fixed_cpt} (top left), the Wang and Aue methods are poorly peaked indicating a change has not beed detected,
while the  Ratio method produces a partial peak indicating it is struggling to detect the change. 
These performances are reflected in the changepoint errors (\ref{fig::fixed_cpt} top right)
In the large $p$ setting (bottom left and right), both the Galeano and Ratio methods are well peaked, 
with the Galeano method producing the lower error. 
The wang method appears to partially locate the change but appears to be biased to the right and 
thus fails to localize the change.
It is unclear what causes this bias. 
Finally, we note that the results in this setting differ from what we would expect based on Section \ref{sec::multiple_cpts_sims}.
In particular, in that experiment we found that the Ratio method completely outperformed the Galeano approach.
Again the reason for this disparity is unclear, 
however given that a greater range of changes are considered in the multiple change setting they may indicate that 
the Galeano approach works better on certain types of changes.

\subsection{Impact of Mean Centering}\label{app:mean_center}
The Ratio method assumes that data has mean zero ( or equivalently that the mean is known).
This is unlikely to be true in practice and some mean centering will typically be required.
When faced with data with an unknown stationary mean, 
we recommend centering the data by subtracting the sample mean. 
To evaluate the impact of this approach, 
we applied the Ratio method to datasets with known and unknown means,
under both the null and alternate hypothesis.
In particular, we consider two sets of dimensions, namely $n=500,p=15$ and $n=2500,p=100$.
For each dimension set and hypothesis, we generated 1000 datasets and applied the Ratio method 
(note under then alternative $\delta=1.15$).
Results from this analysis are shown in Table \ref{tab::mean_centering}.
We can see that the mean centering, has a very marginal impact on the performance of the Ratio method.
\begin{table}[ht]
    \centering
    \begin{tabular}{rrrllrr}
        \hline
        & n & p & hypothesis & mean & TPR & Change Estimate \\
        \hline
        1 & 500.00 & 15.00 & Alt & Mean Known & 0.92 & 222.82 \\
        2 & 500.00 & 15.00 & Alt & Mean Unknown & 0.94 & 225.70 \\
        3 & 500.00 & 15.00 & Null & Mean Known & 0.04 & 238.88 \\
        4 & 500.00 & 15.00 & Null & Mean Unknown & 0.02 & 240.38 \\
        5 & 2500.00 & 100.00 & Alt & Mean Known & 1.00 & 1247.10 \\
        6 & 2500.00 & 100.00 & Alt & Mean Unknown & 1.00 & 1247.14 \\
        7 & 2500.00 & 100.00 & Null & Mean Known & 0.00 & 1159.56 \\
        8 & 2500.00 & 100.00 & Null & Mean Unknown & 0.00 & 1187.74 \\
        \hline
    \end{tabular}
    \caption{Average TPR and estimated change location depending on whether the data was centered by subtracting the sample mean.
    We can see that the centering has little impact on the results.}
    \label{tab::mean_centering}
\end{table}

\input{bias_calculation}

\section{Standard Errors for Error Metrics in Section 5.2}
\begin{table}
\centering
\begin{tabular}{rrrrrrr}
  \hline
  n & p & metric & Aue & Galeano &Ratio &Wang  \\
  \hline
 500  &   3 & FDR & (0.36,0.39) &(0.43,0.49) &(0.25,0.28) &(0.62,0.65) \\
1000 &   3 & FDR & (0.47,0.50) &(0.48,0.54) &(0.27,0.29) &(0.76,0.78) \\
2000 &   3 & FDR & (0.53,0.56) &(0.51,0.56) &(0.28,0.31) &(0.85,0.86) \\
5000 &   3 & FDR & (0.57,0.60) &(0.55,0.59) &(0.28,0.30) &(0.89,0.90) \\
500  &  10 & FDR & (0.24,0.27) &(0.43,0.49) &(0.15,0.17) &(0.26,0.29) \\
1000 &  10 & FDR & (0.30,0.33) &(0.45,0.50) &(0.12,0.13) &(0.41,0.44) \\
2000 &  10 & FDR & (0.33,0.35) &(0.49,0.53) &(0.09,0.10) &(0.52,0.55) \\
5000 &  10 & FDR & (0.36,0.38) &(0.55,0.59) &(0.08,0.09) &(0.60,0.63) \\
2000 &  30 & FDR &   &(0.45,0.49) & (0.01,0.02)&(0.29,0.33) \\
5000 &  30 & FDR &   &(0.51,0.54) & (0.01,0.02)&(0.31,0.34) \\
5000 & 100 & FDR &   &(0.49,0.54) & (0.00,0.00)&(0.41,0.48) \\
500  &   3 & MAE & (25.11,26.99)& (36.45,39.52) & (26.87,28.19)& (41.44,44.20) \\
1000 &   3 & MAE & (17.74,19.13)& (26.73,29.15) & (15.79,16.69)& (35.60,38.02) \\
2000 &   3 & MAE & (12.28,13.28)& (20.35,22.57) & (7.89 ,8.43 )& (31.03,33.24) \\
5000 &   3 & MAE & (7.60 ,8.29 )& (13.15,14.78) & (2.91 ,3.14 )& (22.97,24.72) \\
500  &  10 & MAE & (290.97,317.03)& (517.79,566.01) & (313.69,331.40) & (304.50,329.10) \\
1000 &  10 & MAE & (160.52,174.66)& (390.32,436.14) & (136.61,146.21) & (234.69,252.70) \\
2000 &  10 & MAE & (95.00 ,104.04)& (275.16,309.84) & (49.71 ,54.01 ) & (178.63,193.88) \\
5000 &  10 & MAE & (55.19 ,60.98 )& (174.30,203.93) & (16.88 ,18.35 ) & (117.23,129.46) \\
2000 &  30 & MAE &    & (1491.76,1609.68) & (139.58,147.44) & (1064.90,1127.73) \\
5000 &  30 & MAE &    & (871.37 ,955.54 ) & (49.57 ,52.62 ) & (501.79,544.71)\\
5000 & 100 & MAE &    & (5633.54,6305.87) & (194.38,217.56) & (7614.43,8175.72) \\
500  &   3 & TDR & (0.53,0.57) & (0.19,0.22) & (0.34,0.36) & (0.49,0.52) \\
1000 &   3 & TDR & (0.57,0.60) & (0.22,0.24) & (0.44,0.46) & (0.46,0.50) \\
2000 &   3 & TDR & (0.62,0.65) & (0.26,0.29) & (0.56,0.58) & (0.42,0.46) \\
5000 &   3 & TDR & (0.64,0.67) & (0.30,0.32) & (0.65,0.67) & (0.43,0.47) \\
500  &  10 & TDR & (0.54,0.57) & (0.20,0.23) & (0.51,0.53) & (0.49,0.52) \\
1000 &  10 & TDR & (0.67,0.70) & (0.24,0.27) & (0.70,0.72) & (0.53,0.56) \\
2000 &  10 & TDR & (0.75,0.78) & (0.28,0.31) & (0.84,0.85) & (0.56,0.59) \\
5000 &  10 & TDR & (0.79,0.81) & (0.31,0.34) & (0.89,0.90) & (0.60,0.63) \\
2000 &  30 & TDR &    & (0.34,0.37) & (0.98,0.99) & (0.44,0.47) \\
5000 &  30 & TDR &    & (0.38,0.41) & (0.98,0.99) & (0.54,0.57) \\
5000 & 100 & TDR &    & (0.45,0.50) & (1.00,1.00) & (0.30,0.34) \\
    \hline
\end{tabular}
\caption{95\% confidence intervals for error metric results provided in Table 5.2. Note incorporating the standard errors does not change the results.}
\label{tab:se_51}
\end{table}

\begin{table}
    \centering
        \begin{tabular}{rrrrrrr}
    \hline
             n & p & metric & Aue & Galeano & Ratio & Wang   \\
    \hline
    500 &   3 & FDR &  (.29,.30) & (.93,.95)& (.09,.10) & (.65,.67) \\
    1000 &   3 & FDR &(.40,.41) & (.80,.82)& (.12,.13) & (.80,.81) \\
    2000 &   3 & FDR &(.46,.47) & (.66,.69)& (.15,.16) & (.88,.88) \\
    5000 &   3 & FDR &(.51,.52) & (.50,.53)& (.17,.18) & (.92,.92) \\
    500 &   10 & FDR & (.35,.37) & (1,1)    & (.29,.31) & (.55,.58) \\
    1000 &  10 & FDR &(.44,.46) & (1,1)    & (.24,.26) & (.72,.74) \\
    2000 &  10 & FDR &(.47,.49) & (.99,.99)& (.23,.25) & (.79,.80) \\
    5000 &  10 & FDR &(.47,.49) & (.94,.96)& (.18,.19) & (.80,.82) \\
    2000 &  30 & FDR &  & (0.95, 0.97) & (0.02, 0.03) & (.83,.85) \\
    5000 &  30 & FDR &  & (0.86, 0.88) & (0.02, 0.02) & (.83,.85) \\
    5000 & 100 & FDR &          & (.61,.66)& (0,0) & (.96, .98) \\
    500  &   3 & MAE & (2.87 ,21.54) & (4.02,4.92) & (17.05, 17.44) & (41.29, 42.22) \\
    1000 &   3 & MAE & (16.1 ,16.67)& (31.93, 32.74) &(1.00, 1.28) & (36.9,37.74)\\
    2000 &   3 & MAE & (11.9 ,12.36)& (25.45, 26.22) &(5.46, 5.65) & (3.24,3.9)\\
    5000 &   3 & MAE & (7.40 ,7.69 )& (19.19, 19.83) &(2.37, 2.47) & (21.9,22.5)\\
    500 &  10  & MAE & (225.2,228.3)& (327.9, 330.2) & (248.7, 251) &  (264.1, 266.9) \\
    1000 &  10 & MAE & (139.7,142.4)& (252.1, 254.2) & (145.9, 147.9) &  (20.6, 203) \\
    2000 &  10 & MAE & (89.9,92.19)& (205.1, 207.4) & (75.14,76.55) & (146.1, 148.2) \\
    5000 &  10 & MAE & (5.13,51.82)& (164.7, 167.1) & (24.21,24.87) & (89.55, 91.25) \\
    2000 &  30 & MAE &    & (1326,1333)  & (147.5, 152.9)& (1124, 1132)\\
    5000 &  30 & MAE &    & (92.37, 93.17) & (44.10, 45.64) & (668, 675.8) \\
    5000 & 100 & MAE &    & (6983, 7089) & (186.6,199.9) & (7766, 7825) \\
    500 &   3  & TDR &(.75,.76)&(.01,.02) &(.62,.63) &(.40,.41) \\
    1000 &   3 & TDR &(.76,.78)&(.04,.05) &(.69,.70) &(.35,.36) \\
    2000 &   3 & TDR &(.76,.78)&(.08,.09) &(.75,.76) &(.29,.30) \\
    5000 &   3 & TDR &(.76,.77)&(.15,.16) &(.79,.80) &(.27,.29) \\
    500 &  10  & TDR &(.34,.35)&(.00,.00) &(.26,.27) &(.18,.19) \\
    1000 &  10 & TDR &(.45,.47)&(.00,.00) &(.37,.38) &(.13,.14) \\
    2000 &  10 & TDR &(.53,.55)&(.00,.00) &(.50,.51) &(.12,.13) \\
    5000 &  10 & TDR &(.63,.64)&(.01,.01) &(.68,.69) &(.14,.15) \\
    2000 &  30 & TDR &  & (.01,.01) &(.95 ,.95) & (.04,.05) \\
    5000 &  30 & TDR &  & (.03,.04) &(.97 ,.98) & (.05,.06) \\
    5000 & 100 & TDR &  & (.10,.12) &(1, 1) & (.01,.01) \\
    \hline
\end{tabular}
\caption{95\% confidence intervals for error metric results provided in Table 5.2. Note incorporating the standard errors does not change the results.}
\label{tab:se_52}
\end{table}

\begin{table}[]
    \begin{tabular}{||l|l|l|l|l|l||}
                \hline
            \diagbox{n}{p} & Metric   & 500 & 1000 & 2000 & 5000 \\
                \hline
            10 & FPR & .06 & .039 & .044 & .028 \\
               & TPR & .344 & .701 & .996    & 1 \\
            \hline
            50 & FPR & 0   & .001 & 0    & .001 \\
               & TPR & .045 & .782 & 1    & 1 \\
            \hline
            100& FPR &     & 0    & 0    & .002 \\
               & TPR &     & .502    & 1    & 1 \\
            \hline
    \end{tabular}
    \caption{Detection rates for the Ratio method for data with different dimensions.}
    \label{tab::FPR_TPR}
\end{table}



\end{document}

%% file: ST-input2.tex
\usepackage{amsfonts,amsmath,amssymb,mathrsfs,amsthm}									
\usepackage{graphicx}										
\usepackage[nodayofweek]{datetime}			
\usepackage[ruled, vlined]{algorithm2e}
\usepackage{algpseudocode}
\SetKwInOut{Input}{Input}\SetKwInOut{Output}{Output}

\usepackage{amsmath}
\usepackage{amssymb}
\usepackage{enumerate}
\usepackage{hyperref}
\usepackage{mathtools}
\usepackage{amsthm}
\usepackage{bm}
\usepackage{enumitem}
\usepackage{rotating}
\DeclarePairedDelimiter\abs{\lvert}{\rvert}
\DeclarePairedDelimiter\norm{\lVert}{\rVert}

\newtheorem{definition}{Definition}[section]
\newtheorem{theorem}[definition]{Theorem}
\newtheorem{lemma}[definition]{Lemma}
\newtheorem{proposition}[definition]{Proposition}

\newtheorem{assumption}{Assumption}[section]

\newtheoremstyle{mystyle}  
  {}   				
  {}   				
  {}  				
  {0pt}       		
  {\bfseries}	
  {:}        
  {5pt plus 1pt minus 1pt}				  
  {}          

\theoremstyle{mystyle}
\newtheorem*{eg}{Working Example}

%% file: bias_calculation.tex
\section{Auxillary Results}
    The results in this section are required for the proof of Theorem \ref{thm::bias_var}.
\begin{lemma}
    \label{thm::int1}
    Let $\bm{\gamma} := (\gamma_1, \gamma_2)$ and  $f_1$ be the real valued function
    $$ f_1(x) := (1-x)^2.$$
    Then 
    $$  \underset{r\downarrow 1}{\lim} \frac{1}{4\pi i}
            \oint_{|z|=1} f\left(\frac{|1+h\xi|^2}{(1-\gamma_2)^2} \right)
    \left[\frac{1}{\xi-r^{-1}} + \frac{1}{\xi + r^{-1}} -  
    \frac{2}{\xi + \frac{\gamma_2}{h}}\right] d\xi = 
    2K_3\left(1 -  \frac{\gamma_2^2}{h^2}\right) + \frac{2K_2\gamma_2}{h} $$
    where 
    \begin{align*}
    K_2 =  \frac{2h(1+h^2)}{(1-\gamma_2)^4} -\frac{2h}{(1-\gamma_2)^2},
    \hspace{2mm}
        K_3 = \frac{h^2}{(1-\gamma_2)^4}. 
    \end{align*}
\end{lemma}
\begin{proof}
Firstly we have that  
\begin{align*}
    f_1\left(\frac{|1 + h\xi|^2}{(1-\gamma_2)^2}\right) &= \left( 1 - \frac{(1+h\xi)(1+h\bar{\xi})}{
        (1-\gamma_2)^2}\right)^2 \\
    &= 1 - 2\frac{(1+h\xi)(1+h\bar{\xi})}{(1-\gamma_2)^2} + \frac{(1+h\xi)^2(1+h\bar{\xi})^2}{
        (1-\gamma_2)^4}\\
    &= 1 - 2\frac{1 + h\xi + h\bar{\xi} + h}{(1-\gamma_2)^2} \\
    &+ 
    \frac{(1+4h^2 + h^4) + 2h(1+h^2)\xi + 2h(1+h^2)\bar{\xi} 
    + h^2\xi^2 + h^2\bar{\xi}^2}{(1-\gamma_2)^4}\\
    &= \left(1 - 2\frac{1+h}{(1-\gamma_2)^2} +\frac{1 + 4h^2 + h^4}{(1-\gamma_2)^4}\right) + 
        \left(\frac{2h(1+h^2)}{(1-\gamma_2)^4} -\frac{2h}{(1-\gamma_2)^2}\right)\xi + \\ 
        &\left(\frac{2h(1+h^2)}{(1-\gamma_2)^4} -\frac{2h}{(1-\gamma_2)^2}\right)\bar{\xi} + 
        \left(\frac{h^2}{(1-\gamma_2)^4}\right)\xi^2 + 
        \left(\frac{h^2}{(1-\gamma_2)^4}\right)\bar{\xi}^2 \\ 
    &=K_1 + K_2\xi + K_2\bar{\xi} + K_3\xi^2 + K_3\bar{\xi}^2
\end{align*}

Then 
\begin{align*}
\frac{1}{2\pi i}\oint_{|\xi|=1}
    f_1\left(\frac{|1 + h\xi|^2}{(1-\gamma_2)^2}\right)\left[\frac{1}{\xi-r^{-1}} + \frac{1}{\xi + r^{-1}} -  \frac{2}{\xi + \frac{\gamma_2}{h}}\right] \\
    =  
    \frac{1}{2\pi i}\oint_{|\xi|=1}
    \left(K_1 + K_2\xi + K_3\bar{\xi} + K_4\xi^2 + K_5\bar{\xi}^2\right)\left[\frac{1}{\xi-r^{-1}} + \frac{1}{\xi + r^{-1}} -  \frac{2}{\xi + \frac{\gamma_2}{h}}\right]
\end{align*}
    where $\bar{x}$ is the conjugate of $x$.
    By linearity of the integral we can handle each term separately.
    We can now evaluate the integral using the Cauchy Residue theorem.
    Note the first term is a constant function with respect to $\xi$ and thus cancels out.
    Then 
    \begin{align*}
        \frac{1}{2\pi i}\oint_{|\xi|=1} K_2\xi\left[\frac{1}{\xi-r^{-1}} + \frac{1}{\xi + r^{-1}} -  \frac{2}{\xi + \frac{\gamma_2}{h}}\right] 
        = K_2\left(r^{-1} - r^{-1} + \frac{2\gamma_2}{h}\right) = \frac{2K_2\gamma_2}{h}\\
        \frac{1}{2\pi i}\oint_{|\xi|=1} K_3\xi^2\left[\frac{1}{\xi-r^{-1}} + \frac{1}{\xi + r^{-1}} -  \frac{2}{\xi + \frac{\gamma_2}{h}}\right] = K_3(r^{-2} + r^{-2} -2\frac{\gamma_2^2}{h^2}) = 2K_3\left(1 -  \frac{\gamma_2^2}{h^2}\right)\\
        \frac{1}{2\pi i}\oint_{|\xi|=1} \frac{K_2}{\xi}\left[\frac{1}{\xi-r^{-1}} + \frac{1}{\xi + r^{-1}} -  \frac{2}{\xi + \frac{\gamma_2}{h}}\right] 
        =  K_2(-r  + r -\frac{2h}{\gamma_2} + r - r + \frac{2h}{\gamma_2}) = 0 \\
        \frac{1}{2\pi i}\oint_{|\xi|=1} \frac{K_3}{\xi^2}\left[\frac{1}{\xi-r^{-1}} + \frac{1}{\xi + r^{-1}} -  \frac{2}{\xi + \frac{\gamma_2}{h}}\right] 
        =  K_3\left(-r^2  - r^2 + \frac{2h^2}{\gamma_2^2} + r^2 + r^2 - \frac{2h^2}{\gamma_2^2}\right) = 0 \\
    \end{align*}
    Summing these gives 
    \begin{align}
    2K_3\left(1 -  \frac{\gamma_2^2}{h^2}\right) + \frac{2K_2\gamma_2}{h}.
    \end{align}

\end{proof}

\begin{lemma}
    \label{thm::int2}
    Let $\bm{\gamma} := (\gamma_1, \gamma_2)$ and  $f_1$ be the real valued function
    $$ f_1(x) := (1-x)^2.$$
    Then     
    $$-\underset{r\downarrow 1}{\lim} 
    \frac{2}{4\pi^2 } \oint_{|\xi_1|=1} \oint_{|\xi_2| = 1}
 \frac{1}{(\xi_1-r\xi_2)^2} f_1\left(\frac{|1+h\xi_1|^2}{(1-\gamma_2)^2} \right) 
 f_1\left(\frac{|1+h\xi_2|^2}{(1-\gamma_2)^2} \right) 
 d\xi_2 d\xi_1 = K_2^2 + 2K_3^2$$
 where 
    \begin{align*}
    K_2 =  \frac{2h(1+h^2)}{(1-\gamma_2)^4} -\frac{2h}{(1-\gamma_2)^2},
    \hspace{2mm}
        K_3 = \frac{h^2}{(1-\gamma_2)^4}. 
    \end{align*}
\end{lemma}
\begin{proof}
    Using a similar strategy to the Lemma \ref{thm::int1} we have that 
\begin{align*} 
    -&\frac{1}{4\pi^2}\oint_{|\xi_1|=1} \oint_{|\xi_2|=1} \frac{f_1\left(\frac{|1+h\xi_1|^2}{(1-\gamma_2)^2} \right) 
 f_1\left(\frac{|1+h\xi_2|^2}{(1-\gamma_2)^2} \right)  
}
{(\xi_1 -r\xi_2)^2 } d\xi_1 d\xi_2 \\
    =& -\frac{1}{4\pi^2}\oint_{|\xi_2|=1} f_1\left(\frac{|1+h\xi_2|^2}{(1-\gamma_2)^2} \right)\oint_{|\xi_1|=1} \frac{\left(K_1 + K_2\xi_1 + K_2\xi_1^{-1} + K_3\xi_1^{2} + K_3\xi_1^{-2}\right)
}
{(\xi_1 -r\xi_2)^2 } d\xi_1 d\xi_2 \\
    =&   -\frac{2\pi i }{4\pi^2}\oint_{|\xi_2|=1} 
    f_1\left(\frac{|1+h\xi_2|^2}{(1-\gamma_2)^2} \right)\left(\frac{K_2}{r^2\xi_2^2} + 
    \frac{2K_3}{r^3\xi_2^3}\right)
 d\xi_2\\ 
    =&   -\frac{2\pi i }{4\pi^2}\oint_{|\xi_2|=1} 
\left(K_1 + K_2\xi_2 + K_2\xi_2^{-1} + K_3\xi_2^{2} + K_3\xi_2^{-2}\right)
    \left(\frac{K_2}{r^2\xi_2^2} + 
    \frac{2K_3}{r^3\xi_2^3}\right)
 d\xi_2 \\ 
    =&   -\frac{2\pi i }{4\pi^2}\oint_{|\xi_2|=1} 
    \left(K_1 + K_2\xi_2 + K_3\xi_2^{2} \right) 
    \left(\frac{K_2}{r^2\xi_2^2} + 
    \frac{2K_3}{r^3\xi_2^3}\right)+
    \left(K_2\xi_2^{-1}  + K_3\xi_2^{-2} \right)  
    \left(\frac{K_2}{r^2\xi_2^2} + 
    \frac{2K_3}{r^3\xi_2^3}\right) d\xi_2 \\
    =&   -\frac{2\pi i }{4\pi^2}\oint_{|\xi_2|=1} 
    \left(K_1 + K_2\xi_2 + K_3\xi_2^{2} \right) 
    \left(\frac{K_2}{r^2\xi_2^2} + 
    \frac{2K_3}{r^3\xi_2^3}\right)+
    \left(\frac{K_2^2}{r^2\xi_2^3} + 
        \frac{2K_2K_3}{r^2\xi_2^4} + 
        \frac{K_2K_3}{r^2\xi_2^4} + 
        \frac{2K_2K_3}{r^2\xi_2^5}
    \right) d\xi_2\\ 
 \end{align*}
    Now by the Cauchy Residue Theorem, we have that
    $$\oint_{|\xi_2|=1}\frac{K_2^2}{r^2\xi_2^3} + 
        \frac{2K_2K_3}{r^2\xi_2^4} + 
        \frac{K_2K_3}{r^2\xi_2^4} + 
        \frac{2K_2K_3}{r^2\xi_2^5}d\xi_2 = 0 \text{ and } 
    \oint_{|\xi_2|=1}\left(\frac{K_2}{r^2\xi_2^2} + 
    \frac{2K_3}{r^3\xi_2^3}\right)d\xi_2 = 0,
        $$
    as these expressions can be written as a constant function times a pole of order higher than two.
    Now
    \begin{align*}
        -&\frac{2\pi i }{4\pi^2}\oint_{|\xi_2|=1} 
    \left(K_2\xi_2 + K_3\xi_2^{2} \right) 
    \left(\frac{K_2}{r^2\xi_2^2} + 
    \frac{2K_3}{r^3\xi_2^3}\right)
    d\xi_2 \\ 
        =&\frac{2\pi i }{4\pi^2}\oint_{|\xi_2|=1} 
        \left(\frac{K_2^2}{r^2\xi_2}  + \frac{2K_3^2}{r^3\xi_2}\right)d\xi_2 + 
        \frac{2\pi i }{4\pi^2}\oint_{|\xi_2|=1}
        \left(\frac{2K_2K_3}{r^3\xi_2^2}  + \frac{K_2K_3}{r^2}\right)
    d\xi_2 \\
        =&\frac{2\pi i }{4\pi^2}\oint_{|\xi_2|=1} 
        \left(\frac{K_2^2}{r^2\xi_2}  + \frac{2K_3^2}{r^3\xi_2}\right)d\xi_2  
        =\frac{K_2^2}{r^2} +\frac{2K_3^2}{r^3}
    \end{align*}
    Then taking the limit as $r\downarrow 1$ completes the proof.
\end{proof}

\begin{lemma}
    \label{thm::int3}
    Let $\bm{\gamma} := (\gamma_1, \gamma_2)$ and  $f_1, f_2$ be the real valued function
    $$ f_1(x) := (1-x)^2 \text{ and } f_2(x) := (1 - \frac{1}{x})^2.$$
        Then     
    \begin{align*}
        -\underset{r\downarrow 1}{\lim} 
    \frac{1}{4\pi^2 } \oint_{|\xi_1|=1} \oint_{|\xi_2| = 1}
 \frac{1}{(\xi_1-r\xi_2)^2} f_1\left(\frac{|1+h\xi_1|^2}{(1-\gamma_2)^2} \right) 
 f_2\left(\frac{|1+h\xi_2|^2}{(1-\gamma_2)^2} \right) 
 d\xi_2 d\xi_1 =  \\
    \frac{J_1K_2}{h} + 
    \frac{J_1K_2}{h(h^2-1)} + 
    \frac{-J_1K_3(h^2+1)}{h^2} + 
    \frac{-J_1K_3}{h^2(h^2-1)} + \\
    \frac{J_2K_2 2h}{(h^2-1)^3} + 
    \frac{J_2K_3}{h^2} + 
    \frac{J_2K_3(1-3h^2))}{h^2(h^2-1)^3}  
    \end{align*}
 where 
    \begin{align*}
    K_2 =  \frac{2h(1+h^2)}{(1-\gamma_2)^4} -\frac{2h}{(1-\gamma_2)^2},
    \hspace{2mm}
        K_3 = \frac{h^2}{(1-\gamma_2)^4}  \\
    J_1 = - 2(1-\gamma_2)^2 \text{ and } J_2 = (1-\gamma_2)^4.
    \end{align*}
\end{lemma}
\begin{proof}
    Firstly we have that, 
    \begin{align*}
        f_2\left(\frac{|1 + h\xi|^2}{(1-\gamma_2)^2}\right) &=
        \left(1 - \frac{(1-\gamma_2)^2}{(1+h\xi_2)(1+h\bar{\xi}_2)} \right)^2 \\
        &= 1 - 2\frac{(1-\gamma_2)^2}{(1+h\xi_2)(1+h\bar{\xi}_2)}  +
        \frac{(1-\gamma_2)^4}{(1+h\xi_2)^2(1+h\bar{\xi}_2)^2} \\
        &= 1 + \frac{J_1}{(1+h\xi_2)(1+h\bar{\xi}_2)}  +
        \frac{J_2}{(1+h\xi_2)^2(1+h\bar{\xi}_2)^2} \\
        &= 1 + \frac{J_1\xi_2}{(1+h\xi_2)(\xi_2+h)}  +
        \frac{J_2\xi_2^2}{(1+h\xi_2)^2(\xi_2+h)^2}
    \end{align*}
    Using the same constants as in Lemmas \ref{thm::int1} and \ref{thm::int2} we have the following, 
\begin{align*}
    -\frac{1}{4\pi^2}&\oint_{|\xi_1|=1} \oint_{|\xi_2|=1} 
    \frac{
        \left(K_1 + K_2\xi_1 + K_2\xi_1^{-1} + K_3\xi_1^{2} + K_3\xi_1^{-2}\right)
        \left(
            1+ \frac{J_1 \xi_2}{(1+h\xi_2)(\xi_2+h)} + 
            \frac{J_2\xi_2^2}{(1+h\xi_2)^2(\xi_2+h)^2}
        \right)
    }{
        (\xi_1 -r\xi_2)^2 
    } 
    d\xi_1 d\xi_2 
    \\ 
    =-&\frac{1}{4\pi^2}\oint_{|\xi_1|=1} \oint_{|\xi_2|=1} 
    \frac{
        \left(K_1 + K_2\xi_1 + K_2\xi_1^{-1} + K_3\xi_1^{2} + K_3\xi_1^{-2}\right)
    }{
        (\xi_1 -r\xi_2)^2 
    } 
    d\xi_1
    \\
    &\times \left(
        1+ \frac{J_1 \xi_2}{(1+h\xi_2)(\xi_2+h)} + 
        \frac{J_2\xi_2^2}{(1+h\xi_2)^2(\xi_2+h)^2}
    \right)
d\xi_2 
    \\
    = -&\frac{2 \pi i}{4\pi^2} \oint_{|\xi_2|=1} 
    \left(\frac{K_2}{r^2\xi_2^2} + \frac{K_3}{r^3 \xi_2^3}\right) 
    \left(
        1+ \frac{J_1 \xi_2}{(1+h\xi_2)(\xi_2+h)} + 
        \frac{J_2\xi_2^2}{(1+h\xi_2)^2(\xi_2+h)^2}
    \right)
    d\xi_2 
    \\
    = -&\frac{2 \pi i}{4\pi^2} \oint_{|\xi_2|=1} 
    \frac{K_2}{r^2\xi_2^2} + \frac{K_3}{r^3 \xi_2^3} +  
    \frac{J_1 K_2}{r^2\xi_2(1+h\xi_2)(\xi_2+h)} + 
    \frac{J_1 K_3}{\xi_2^2(1+h\xi_2)(\xi_2+h)} + 
    \\
     & \frac{J_2K_2}{r^2(1+h\xi_2)^2(\xi_2+h)^2} + 
      \frac{J_2K_3}{r^2\xi(1+h\xi_2)^2(\xi_2+h)^2} d\xi_2
    \\= -&\frac{2 \pi i}{4\pi^2} \oint_{|\xi_2|=1} 
    \frac{J_1 K_2}{r^2\xi_2(1+h\xi_2)(\xi_2+h)} + 
    \frac{J_1 K_3}{\xi_2^2(1+h\xi_2)(\xi_2+h)} + 
    \\
     & \frac{J_2K_2}{r^2(1+h\xi_2)^2(\xi_2+h)^2} + 
      \frac{J_2K_3}{r^2\xi_2(1+h\xi_2)^2(\xi_2+h)^2} d\xi_2
    \\
    = -&\frac{2 \pi i}{4\pi^2} \oint_{|\xi_2|=1}
    ((i) + (ii) + (iii) + (iv)) d\xi_2
\end{align*}
These values can be calculated using the residue theorem.

\begin{tabular}{c | c | c | c | c}
    Term & (i) & (ii) & (iii) & (iv) \\
    Residue Locations & 0, -h &  0, -h &  -h &  0, -h  \\ 
    Orders & 1,1 & 2, 1 & 2 & 1, 2
\end{tabular}

Then the integral is given by the following,
\begin{align}
    \frac{J_1K_2}{h} + 
    \frac{J_1K_2}{h(h^2-1)} + 
    \frac{-J_1K_3(h^2+1)}{h^2} + 
    \frac{-J_1K_3}{h^2(h^2-1)} + \\
    \frac{J_2K_2 2h}{(h^2-1)^3} + 
    \frac{J_2K_3}{h^2} + 
    \frac{J_2K_3(1-3h^2)}{h^2(h^2-1)^3}. 
\end{align}

\end{proof}

\section{Proof of Main Results}
\label{sec::cov_results}
In this section, we provide proofs for the main results in the chapter.
\begin{proof}[Proof of Proposition \ref{thm::invariance}]
        Firstly 
        $$ R(\Sigma_1, \Sigma_2) = {(\Sigma_2)}^{-1} \Sigma_1 = 
        {\left( {( \Sigma_1)}^{-1}  \Sigma_2\right)}^{-1} = (R(\Sigma_2, \Sigma_1))^{-1},$$
        which implies that
        $$ \lambda_j(R(\Sigma_1, \Sigma_2)) = \lambda_j^{-1}(R(\Sigma_2, \Sigma_1)) $$
        Then
        \begin{align*}
            T(\Sigma_1, \Sigma_2) &= \sum_{j=1}^p \left(1 - \lambda_j(R(\Sigma_1,\Sigma_2))\right)^2 
            + \left(1 - \lambda_j^{-1}(R(\Sigma_1,\Sigma_2))\right)^2  \\
            &= \sum_{j=1}^p \left(1 - \lambda_j^{-1}(R(\Sigma_2,\Sigma_1))\right)^2 
            + \left(1 - \lambda_j(R(\Sigma_2,\Sigma_1))\right)^2. 
        \end{align*}
        Now the final term is the definition of the test statistic $T(\Sigma_2, \Sigma_1)$.
        Thus 
        $$ T(\Sigma_1, \Sigma_2) = \sum_{j=1}^p \left(1 - \lambda_j^{-1}(R(\Sigma_2,\Sigma_1))\right)^2 
            + \left(1 - \lambda_j(R(\Sigma_2,\Sigma_1))\right)^2 = T(\Sigma_2, \Sigma_1)$$
        proving symmetry.

        Secondly 
        $$ R(\Sigma_1, \Sigma_2) = {(\Sigma_2)}^{-1} \Sigma_1 = 
        {\left( {( \Sigma_1)}^{-1}  \Sigma_2\right)}^{-1} =  
        {\left( \Sigma_2{( \Sigma_1)}^{-1}  \right)}^{-T}
        (R(\Sigma_2, \Sigma_1))^{-T},$$
        which implies that
        $$ \lambda_j(R(\Sigma_1, \Sigma_2)) = \lambda_j^{-1}(R(\Sigma_1^{-1}, \Sigma_1^{-1})) $$
        which in turn implies
        $$ T(\Sigma_1, \Sigma_2) = T(\Sigma_1^{-1}, \Sigma_2^{-1}).$$

        Finally
        \begin{align*}
            \lambda_j(R(\Sigma_1, \Sigma_2)) = 
             \lambda_j\left(\left(\Sigma_0Z_2^T Z_2 \Sigma_0 \right)^{-1}
             \Sigma_0Z_1^T Z_1 \Sigma_0\right)  \\ 
            = \lambda_j\left(\Sigma_0^{-1}\left(Z_2^T Z_2\right)^{-1} \Sigma_0^{-1} 
              \Sigma_0Z_1^T Z_1 \Sigma_0\right) = \\ 
            = \lambda_j\left(\Sigma_0^{-1}\left(Z_2^T Z_2\right)^{-1} 
            \bm{Z}_1^T Z_1 \Sigma_0\right) = \\ 
            = \lambda_j\left(\left(Z_2^T Z_2\right)^{-1} 
            \bm{Z}_1^T Z_1 \Sigma_0\Sigma_0^{-1}\right) =
            \lambda_j\left(\left(Z_2^T Z_2\right)^{-\frac{1}{2}} 
            \bm{Z}_1^T Z_1 \right) = \lambda_j(R(Z_1^T Z_1, Z_2^TZ_2)) .
        \end{align*} 
        Hence $T(\Sigma_1, \Sigma_2)  = T(Z_1^T Z_1, Z_2^TZ_2)$.
    \end{proof}

    The proof of Theorem \ref{thm::bias_var} requires the application of Theorem 3.1 \cite{zheng2012}.
    For completeness, we state the this result in full below.
    \begin{theorem}{\cite{zheng2012}}
        \label{thm::zheng}
        Let $X \in \mathbb{R}^{n_1 \times p}$ and $Y \in \mathbb{R}^{n_2 \times p}$ be random matrices
        satisfying Assumption \ref{assump1},
        and $f_1, \dots, f_s$ (s is a fixed integer) be functions analytic in an open region in 
        the complex plane containing the interval $[a_{\gamma}, b_{\gamma}]$.
        Then, as $\bm{n} \to \infty$, the random vector
        $$ \left[ \int f_k(x) d G_n(x) \right] 1 \leq k \leq s$$ 
        converges weakly to a Gaussian vector $(X_{f_1}, \dots X_{f_s})$ with means, $\mu_{f_k}$,
        and variances, $\sigma_{f_k}^2$, 
        \begin{align}
            \mathbb{E}_{f_k}(\bm{\gamma}) := \underset{r\downarrow 1}{\lim} \frac{1}{4\pi i}
                \oint_{|z|=1} f\left(\frac{|1+h\xi|^2}{(1-\gamma_2)^2} \right)
        \left[\frac{1}{\xi-r^{-1}} + \frac{1}{\xi + r^{-1}} -  
            \frac{2}{\xi + \frac{\gamma_2}{h}}\right] d\xi  \label{eq::mean}\\ 
            Cov_{f_k, f_j}(\bm{\gamma}) := -\underset{r\downarrow 1}{\lim} \frac{2}{4\pi^2 } \oint_{|\xi_1|=1} 
     \oint_{|\xi_2| = 1}
     \frac{1}{(\xi_1-r\xi_2)^2} f\left(\frac{|1+h\xi_1|^2}{(1-\gamma_2)^2} \right) 
     f\left(\frac{|1+h\xi_2|^2}{(1-\gamma_2)^2} \right) 
            d\xi_2 d\xi_1. \label{eq::var} 
        \end{align}
    \end{theorem}
    We now use the above result, and the results in the previous section to prove the main result of the paper.

\begin{proof}{Proof of Theorem \ref{thm::bias_var}}
    Let $t_1(x) = (1- x)^2$ and $t_2(x) = (1 - \frac{1}{x})^2$. 
    Then by Theorem \ref{thm::zheng} the vector 
    $\bm{t}_n(x) := (\int f_1(x) dF_n(x), \int f_2(x) dF_n(x))$ 
    converges to a Normal vector with mean and covariance given by equations \eqref{eq::mean} and \eqref{eq::var}.
    Now our test statistic (at a single time point) can be expressed as $\bm{1}^T \bm{t}_n(x)$ and thus by the continuous mapping theorem converges weakly to a Normal random variable with moments
    \begin{align} 
        \label{eq::formula}
        \mathbb{E}_{f_1}(\gamma) + \mathbb{E}_{f_2}(\gamma) \text{ and }
        Cov_{f_1, f_1}(\gamma) + 2Cov_{f_1, f_2}(\gamma) + Cov_{f_2, f_2}(\gamma).
    \end{align}
    We also have the following relationship between $t_1$ and $t_2$, 
    $$ t_1(\lambda_j(\bm{\Sigma}_1^{-1} \bm{\Sigma}_2)) =
        (1 - \lambda_j(\bm{\Sigma}_1^{-1} \bm{\Sigma}_2))^2 = 
        (1 - \lambda_j(\bm{\Sigma}_2^{-1} \bm{\Sigma}_1))^2 = 
     t_2(\lambda_j(\bm{\Sigma}_2^{-1} \bm{\Sigma}_1)).$$
     By Theorem \ref{thm::zheng}, the limiting distributions of $f_1$ and $f_2$ depend on 
     $\bm{\gamma}$ which implies that 
     \begin{align}
         \mathbb{E}_{t_1}(\gamma_1, \gamma_2) = \mathbb{E}_{t_2}(\gamma_2, \gamma_1) 
         \text{ and } 
         Cov_{t_1,t_1}^2(\gamma_1, \gamma_2) = Cov_{t_2,t_2}^2(\gamma_2, \gamma_1).
     \end{align}

     By Lemma \ref{thm::int1} we have that 
     $$\mathbb{E}_{t_1}(\gamma) = 2K_{3,1}\left(1 -  \frac{\gamma_2^2}{h^2}\right) + \frac{2K_{2,1}\gamma_2}{h}$$
     where 
     $$ 
    K_{2,1} =  \frac{2h(1+h^2)}{(1-\gamma_2)^4} -\frac{2h}{(1-\gamma_2)^2},
    \hspace{2mm}
        K_{3,1} = \frac{h^2}{(1-\gamma_2)^4}.  
    $$
    By symmetry 
     $$\mathbb{E}_{t_2}(\gamma) = 2K_{3,2}\left(1 -  \frac{\gamma_1^2}{h^2}\right) + \frac{2K_{2,2}\gamma_1}{h}$$
     where
    $$
    K_{2,2} =  \frac{2h(1+h^2)}{(1-\gamma_1)^4} -\frac{2h}{(1-\gamma_1)^2},
    \hspace{2mm}
        K_{3,2} = \frac{h^2}{(1-\gamma_1)^4}.  
        $$
    Combining these values gives the expectation.

    By Lemma \ref{thm::int2} we have that 
    $$Cov_{t_1, t_1}(\gamma) = 2\left( K_{2,1}^2 + 2K_{3,1}^2\right)$$
    and by symmetry we have that  
    $$Cov_{t_2, t_2}(\gamma) = 2\left( K_{2,2}^2 + 2K_{3,2}^2\right).$$
    Finally by Lemma \ref{thm::int3} we have that 
    \begin{align*}
        Cov_{t_1, t_2}(\gamma) = 2\Bigg(\frac{J_1K_{2,1}}{h} + 
        \frac{J_1K_{2,1}}{h(h^2-1)} + 
        \frac{-J_1K_{3,1}(h^2+1)}{h^2} + 
        \frac{-J_1K_{3,1}}{h^2(h^2-1)} + \\
        \frac{J_2K_{2,1} 2h}{(h^2-1)^3} + 
        \frac{J_2K_{3,1}}{h^2} + 
        \frac{J_2K_{3,1}(1-3h^2)}{h^2(h^2-1)^3}\Bigg) 
    \end{align*}
    where 
    $$ 
    J_1 = - 2(1-\gamma_2)^2 \text{ and } J_2 = (1-\gamma_2)^4.$$
    Plugging these values into \eqref{eq::formula} gives the required result.
\end{proof}